\documentclass[preprint,12pt,longnamesfirst]{aastex}

\usepackage{latexsym}

\shorttitle{SNRs in M101}
\shortauthors{Franchetti et al.}

\newcommand {\ha}{H$\alpha$}
\newcommand {\nii}{[\ion{N}{2}]}
\newcommand {\hii}{\ion{H}{2}}

\newcommand {\sii}{[\ion{S}{2}]}

\newcommand {\kms}{km~s$^{-1}$}

\begin{document}

\title{Physical Structure and Nature of Supernova Remnants in M101}

\author{Nicholas A. Franchetti\altaffilmark{1,2}, Robert A. Gruendl\altaffilmark{1}, You-Hua Chu\altaffilmark{1},
Bryan C. Dunne\altaffilmark{1}, Thomas G. Pannuti\altaffilmark{3}, Kip D. Kuntz\altaffilmark{4}, C.-H. Rosie Chen\altaffilmark{5,6}, Caleb K. Grimes\altaffilmark{3}, Tabitha M. Aldridge\altaffilmark{7}}
\altaffiltext{1}{Astronomy Department, University of Illinois, 
        1002 W. Green Street, Urbana, IL 61801, USA;
        franche1@illinois.edu,gruendl@astro.illinois.edu,yhchu@astro.illinois.edu,bdunne@astro.illinois.edu}
\altaffiltext{2}{Department of Aerospace Engineering, University of Illinois, 
        104 South Wright Street, Urbana, IL 61801, USA}
\altaffiltext{3}{Department of Earth and Space Sciences, Space Science Center, Morehead State University,
235 Martindale Drive, Morehead, KY 40351; t.pannuti@moreheadstate.edu,ckgrim01@moreheadstate.edu}
\altaffiltext{4}{The Henry A. Rowland Department of Physics and Astronomy, Johns Hopkins University, 3400 North Charles Street, Baltimore, MD 21218, USA; kuntz@pha.jhu.edu}
\altaffiltext{5}{Department of Astronomy, University of Virginia, Charlottesville, VA 22904, USA}
\altaffiltext{6}{Now at Max Planck Institut f\"ur Radioastronomie, Auf dem H\"ugel 69, D-53121 Bonn, Germany; rchen@mpifr-bonn.mpg.de}
\altaffiltext{7}{Department of Geology and Environmental Geosciences, Northern Illinois University, Davis Hall 312,
Normal Rd., DeKalb, IL 60115; z1611057@students.niu.edu}

\begin{abstract}
Supernova remnant (SNR) candidates in the giant spiral galaxy M101 have been previously identified from ground-based \ha\ and \sii\ images. 
We have used archival \emph{Hubble Space Telescope} (\emph{HST}) \ha\ and broad-band images as well as stellar photometry of 55 SNR candidates to examine their physical structure, interstellar environment, and underlying stellar population. 
We have also obtained high-dispersion echelle spectra to search for shocked high-velocity gas in 18 SNR candidates, and identified X-ray counterparts to SNR candidates using data from archival 
observations made by the \emph{Chandra X-ray Observatory}. Twenty-one of these 55 SNR candidates studied have X-ray counterparts, although one of them is a known ultra-luminous X-ray source. 
The multi-wavelength information has been used to assess the nature of each SNR candidate. We find that within this limited sample, $\sim$16\% are likely remnants of 
Type Ia SNe and $\sim$45\% are remnants of core-collapse SNe. In addition, about $\sim$36\% are large candidates which we suggest are either superbubbles or OB/\hii\ complexes. 
Existing radio observations are not sensitive enough to detect the non-thermal emission from these SNR candidates. Several radio sources are coincident with X-ray sources, but they 
are associated with either giant \hii\ regions in M101 or background galaxies. 
The archival \emph{HST} \ha\ images do not cover the entire galaxy and thus prevents a complete study of M101. 
Furthermore, the lack of  \emph{HST} \sii\ images precludes searches for small SNR candidates which could not be identified by ground-based observations. 
Such high-resolution images are needed in order to obtain a complete census of SNRs in M101 for a comprehensive investigation of the distribution, population, and rates of SNe in this galaxy. 
\end{abstract}


\section{Introduction}
Mixing of supernova (SN) ejecta and the interstellar medium (ISM) occurs at
supernova remnant (SNR) sites, dispersing elements formed by fusion in 
stellar interiors as well as heavier elements created during 
supernova explosions. SNRs not only raise the metal content of the ISM, 
but also heat the ISM by releasing a large amount of kinetic energy. Thus SNRs 
are crucial to the evolution  of the ISM as well galaxies as a whole.

SNe can be generally divided into two categories: those formed 
by high-mass progenitors undergoing core-collapse, and those formed by white 
dwarfs in binary systems \citep{WS88}. The latter, Type Ia SNe, occur when a 
white dwarf accretes material from its binary companion, 
causing the white dwarf to exceed the Chandrasekhar limit. Type Ib, Ic, 
or II SNe are caused by massive stars undergoing core-collapse. 
The two categories of SNe inject different heavy elements into the ISM of 
a galaxy, and therefore have a different impact on galactic chemical evolution. 
SNRs resulting from core-collapse SNe are typically located in a galactic plane, 
while Type Ia SNRs can be found anywhere in a galaxy, even far above 
its galactic plane in the halo. The differing spatial distributions and elemental 
contributions of the different types of SNRs suggest 
that studying the rates and types of SNRs in a galaxy is vital to 
understanding the structure and evolution of its ISM.

It is difficult to obtain a global view of SNRs within the Galaxy, because their distances are not well-known.  
The absence of accurate distances to Galactic SNRs means we cannot study properties such as size and 
luminosity, and prevents us from creating a comprehensive census of the properties and distribution of Galactic 
SNRs. In an external galaxy, all the SNRs are located at approximately the same known distance and we can 
view their relative positions, allowing an accurate determination of the distribution of SNRs within the galaxy 
\citep{MF97,Wetal99,Petal07,Letal10,Petal11}. However, ground-based surveys of extragalactic SNRs outside of the 
Local Group have poor angular resolution, and cannot provide sufficiently detailed images to allow for a 
comprehensive study of SNRs in a galaxy. The \emph{Hubble Space Telescope} (\emph{HST}), with its high angular 
resolution, provides images that are ideal for the study of SNRs in nearby galaxies.  

M101 is a nearly face-on spiral galaxy \citep[$\sim$18$^\circ$ inclination;][]{Betal81} at a distance of 7.4 Mpc 
\citep{Ketal96} and has been observed with the \emph{HST} in \ha\ and continuum bands. 
These \emph{HST} images, in conjunction with available complementary observations, make it possible to 
assess the nature of previously identified SNR candidates.  
Ninety-three SNR candidates in M101 have been identified by \citet{MF97} using ground-based optical images 
and the selection criterion of \sii/\ha\  $\geq$ 0.45. They acknowledged the need for a more detailed study of the 
SNR candidates they identified, especially those with a diameter $\geq$ 100 pc. \emph{HST} images resolved 
these SNR candidates so that detailed morphologies can be examined and accurate sizes can be measured. 
We have found archived \emph{HST} \ha\ images for 55 of the 93 SNR candidates identified by \citet{MF97}, 
and have conducted an analysis of their physical structure and nature. The \emph{HST} images are used to measure 
the sizes of the SNRs and to analyze the underlying stellar population, which may provide clues to help identify the nature of their SN progenitors.

Classical SNRs exhibit three identifying characteristics: high \sii/\ha\ ratio, non-thermal radio spectral index, and bright X-ray emission. 
These characteristics are produced by high-velocity shocks; therefore, high-velocity gas is usually detected in SNRs unless a SNR 
is in a low-density medium such as the interior of a superbubble \citep{C97}. We have thus obtained high-dispersion 
echelle spectra of some M101 SNR candidates to search for high-velocity gas. We have also examined 
\emph{Chandra} X-ray observations of M101 as well as published lists of discrete radio sources in M101 
to search for counterparts to the optically identified SNR 
candidates. This multi-wavelength study helps better assess the true nature of the SNR 
candidates in M101. The method outlined in this study can be used to analyze a complete sample of SNRs in a galaxy to establish the relative 
frequencies and distributions of Type 1a and core-collapse SNRs.

The results of our analysis of previously identified M101 SNR candidates are reported in this paper. Section 2 describes the observations 
used in this study, Section 3 reports the physical properties of the SNR candidates, Section 4 examines their 
underlying stellar population, Section 5 discusses the X-ray and radio observations of M101, and Section 6 assesses the physical 
nature of the SNR candidates. A summary is given in Section 7. 

\section{Observations}
The data sets used in our study include: (1) archived \emph{HST} optical 
images and photometric data, (2) high-dispersion echelle spectra obtained 
with the Kitt Peak National Observatory (KPNO) Mayall 4 m telescope, and (3) \emph{Chandra} X-ray images.

\subsection{Optical Imaging and Photometry}
\emph{HST} Advanced Camera for Surveys, Wide Field Channel (ACS/WFC) and 
Wide Field and Planetary Camera 2 (WFPC2) optical images were obtained 
from the Space Telescope Science Institute Hubble Legacy Archive\footnote{Available at http://hla.stsci.edu/}. A 
detailed list of the archived images used in this study is given in Table 1. 
The WFPC2 images were made through the F656N (\ha) and F606W (Wide $V$) filters. 
The ACS/WFC images were made through the F658N (\ha), 
F435W ($\sim$$B$), F555W ($\sim$$V$), and F814W ($\sim$$I$) filters. 

Fifty-five of the SNR candidates identified by \citet{MF97} had \emph{HST} \ha\ images available 
(see Figure 1 for the coverage of these \emph{HST} \ha\ images). 
These \ha\ images were used to analyze the sizes and morphologies of the SNRs. 
Among the continuum images, we primarily used those in the F435W, F555W, and F814W bands, 
corresponding to $B$, $V$, and $I$, that are commonly used for stellar population analysis. 
Furthermore, the majority of the 55 SNR candidates which had archived \ha\ observations also had images in these continuum bands. 
Only three SNR candidates, MF 25, MF 28, and MF 82, did not have F435W, F555W, and F814W images available. For these three, 
we searched the \emph{HST} archive for observations made through other filters, and
found a WFPC2 F606W image containing MF 25 and MF 28. No such image was available for MF 82. 
The remaining 38 SNR candidates identified by \citet{MF97} either did not have archived 
\emph{HST} \ha\ images available or the archived images had too low a S/N ratio to allow for detailed study, 
thus they are not included in this study. The \ha\ and F555W images of the SNRs are presented in 
Figure 2. The accuracy of the astrometric solutions for these images is typically better than 0\farcs25.

We have made use of the photometric data from the Hubble Legacy Archive. 
These photometric measurements were made with DAOPHOT for point sources in the F435W, 
F555W, and F814W images. These measurements typically have magnitude zero point errors $<$0.1 mag, RMS scatter $<$0.3 mag versus 
published catalogs, and $<$20\%\ artifacts. These photometric data are used to construct color-magnitude 
diagrams (CMDs) for regions encompassing the SNR candidates in order to analyze the 
underlying stellar population. 

\subsection{Spectroscopy}
High-dispersion echelle spectra were obtained with the KPNO Mayall 4 m 
telescope for 18 SNR candidates in two observing 
runs: 1999 July 02 and 2000 April 22-24 (UT). 
Most of these targets were selected because of their large sizes, and the 
presence or absence of high-velocity gas can help us discern between 
bona-fide SNRs and superbubbles. 
Seventeen of these objects 
have available \emph{HST} images. The 79 line mm$^{-1}$ 
echelle grating 79-63 and the 226 line mm$^{-1}$ cross disperser 226-1 
were used in conjunction with the long-focus red camera to achieve a 
reciprocal dispersion of 3.5 \AA\ mm$^{-1}$ at the \ha\ line. The spectra 
were imaged with the 2048 $\times$ 2048 T2KB CCD detector. The 24 $\mu$m 
pixel size corresponds to 0\farcs285 pixel$^{-1}$ along the slit and $\sim$3.7 
\kms\ pixel$^{-1}$ along the dispersion axis at the \ha\ line. 
Each echellogram covers at 
least 4,300-7,000 \AA, which includes the nebular emission lines. Throughout this work, we 
focus on the \ha\ lines as the \nii\ and \sii\ lines are too faint for detailed analysis. 

The journal of observations for the M101 SNR candidates is given in Table 
2. Observations were made using an east-west slit position, with varying 
slit widths. The instrumental FWHM, determined by applying Gaussian 
fits to sky lines, ranges from 12 to 16 \kms. Reduction and analysis of the data was done with 
Interactive Reduction and Analysis Facility (IRAF) software. 

\subsection{Chandra X-ray Observations}
A mega-second \emph{Chandra} X-ray mosaic of M101 obtained and detailed by \citet{KS10} 
was also included in this study to search for X-ray counterparts to each of the 
SNR candidates identified by \citet{MF97}. Since this mosaic covered the entire 
field of M101, we were able to use it to search for X-ray counterparts to all 
93 of the SNR candidates. Small SNRs in M101 will be unresolved by \emph{Chandra's} 
point-spread function and appear as point sources. The point source detection limit of the 
mega-second \emph{Chandra} observation of M101 is $\sim$10$^{36}$ ergs s$^{-1}$ in the 0.5 to 2.0 keV band \citep{KS10}. 
As typical SNRs have luminosities of $\sim$10$^{35}$ to a few times 10$^{36}$ ergs s$^{-1}$ \citep{Metal83}, the bright 
SNRs in M101 should be detectable. 
For M101 SNR candidates with X-ray counterparts, the emission is
generally too weak for a detailed spectral analysis.  On the other hand,
the hardness ratio can often be used to distinguish between sources with
thermal and non-thermal X-ray spectral properties (based primarily on
the absence or presence of hard X-rays).  In Table 3 (Column 10) we
report whether each SNR candidate is coincident with an X-ray counterpart
and note whether the hardness ratios point to a thermal, non-thermal, or
ambiguous origin for the X-ray emission.

\section{Physical Properties of SNR Candidates}
The optically identified SNR candidates in this study are listed in Table 3. The assigned numbers are the same 
as those given by \citet{MF97}, with the SNRs numbered by increasing R.A. 
We have measured accurate positions of the SNRs, and these refined R.A. and Dec. 
are listed in Columns 2 and 3 of Table 3.

\subsection{Sizes of SNR Candidates}
We measured the size of each SNR candidate from its \emph{HST} \ha\ 
image. In Table 3, we list both angular and linear diameters in arcseconds 
and parsecs, respectively, adopting a distance of 7.4 Mpc \citep{Ketal96}, where 1\arcsec\ corresponds to 36 pc. 
Major and minor axes are given for non-spherical SNRs. Linear sizes of the SNR candidates range from $\sim$20 
to 330 pc in diameter.

To compare our size measurements from \emph{HST} images to those measured 
by \citet{MF97}, we converted their published linear sizes to angular sizes using 
their adopted distance of 5.4 Mpc. The comparison of angular sizes is shown in 
Figure 3. To correct for seeing, Matonick and Fesen deconvolved their measured SNR diameters using the average 
FWHM of field stars. This may have resulted in an overcorrection for small objects ($\leq$ 2\arcsec), such as MF 28, MF 75, and MF 90, 
for which Matonick and Fesen reported sizes smaller than 
those measured from \emph{HST} images. The ground-based images 
cannot resolve SNR candidates in complex environments and may have resulted in 
overestimations of sizes in some cases, such as MF 08, MF 12, MF 51, and MF 67. Four SNR candidates, 
MF 19, MF 53, MF 78, and MF 83, are large and ill-defined, thus the size and shape are 
subjective, resulting in a large scatter. Aside from these differences which can be related to  
resolution and sensitivity, the agreement between our measurements is reasonable.

Sixteen of the 55 SNR candidates we studied have diameters greater than 
100 parsecs. Such large sizes are very rare among known SNRs \citep{Wetal99}. 
Some of the large SNR candidates identified by \citet{MF97} may be superbubbles, 
because they are identified solely by \sii/\ha\ ratio, and it has been shown that 
superbubbles can have \sii/\ha\  $\geq$ 0.45 
\citep{Cetal00,L77}. Indeed, several of these large SNR candidates do not have a 
clearly defined shell structure and encompass \hii\ regions and OB associations, as 
discussed in Section 4. 

Only two of the 55 SNR candidates we studied had a diameter $\leq$ 20 pc (0\farcs55). 
The lack of a significant number of small SNRs confirms the existence 
of the selection effect \citet{MF97} suggested in their study. This is to be expected because 
ground-based observations do not allow for detection of objects that are very small 
or very narrow (unresolved) in a galaxy as distant as M101. 
Furthermore, young Type Ia SNRs are dominated by Balmer line emission without forbidden 
lines \citep{Setal91} and young core-collapse SNRs with O-rich ejecta (such as the SNR in NGC 4449) are identified 
by strong oxygen forbidden lines \citep{Betal83}; these young SNRs would have been missed by \citet{MF97} because of the \sii/\ha\ $\geq$ 0.45 selection criterion. 
It is likely that a complete \ha\ and \sii\ \emph{HST} imaging survey of M101 in conjunction with 
\emph{Chandra} X-ray data would detect a significant number of smaller SNR candidates. 
Only with a complete census of SNRs in M101 can we carry out a comprehensive study of the properties and 
spatial distribution of SNRs and their contribution to the evolution of a giant spiral galaxy.

\subsection{Expansion Velocities of SNR candidates}
We have obtained high-dispersion echelle spectra for 18 SNR candidates and used the 
\ha\ line to determine their kinematic properties. 
Measuring a distinct value for the expansion velocity of a SNR is nearly impossible, 
because all sides of the remnant will be expanding at different rates due to 
differences in the density of the surrounding ISM, with the largest expansion 
velocities occurring where the ISM is least dense. Furthermore, because of the large 
distance to M101, the faint high-velocity material in the SNR is more difficult to detect.  

The observed \ha\ wavelengths are converted to heliocentric radial velocities assuming a rest wavelength of 6562.80\AA.
In the long-slit echelle 
spectra, shown in Figure 4, the unshocked ISM will appear narrow and have a nearly constant velocity along the slit, 
while the SNR-shocked material will appear at higher velocities. We adopt the velocity of the 
unshocked ISM as the systemic velocity ($V_{\rm sys}$) and measure the largest blue-shifted and 
red-shifted velocity offsets, $\Delta$$V_{\rm blue}$ and $\Delta$$V_{\rm red}$, caused by the 
expansion of the approaching and receding sides of the SNR candidate, respectively.  
In the cases where no narrow ISM component is obvious, we adopt the centroid of the \ha\ line 
as the systemic velocity. The results are listed in Table 4. $V_{\rm sys}$ ranges from 181 to 264 
\kms\ because of the galactic rotation. $\Delta$$V_{\rm red}$ ranges from $<$34 to 194 \kms, 
and $\Delta$$V_{\rm blue}$ ranges from $-$39 to $-$161 \kms. The uncertainty in these velocity measurements is $\sim$10 to 20 \kms. 
Additional uncertainties in these measurements are caused by the limited slit coverage of the objects and the fact that only the line-of-sight 
component of expansion can be measured. These limitations may prevent the detection of the highest velocity offset. 

We may compare the expansion velocities of the M101 SNR candidates with those in the 
Magellanic Clouds (MCs). \citet{CK88a} and \citet{C97} found that all confirmed SNRs in the MCs have 
$\Delta$$V_{\rm red}$ and/or $-\Delta$$V_{\rm blue}$ greater than 100 \kms, while the superbubble 
N70 has a much lower expansion velocity. The $\Delta$$V_{\rm red}$ and $-\Delta$$V_{\rm blue}$ 
for M101 SNR candidates in Table 4 are mostly smaller than 100 \kms. These differences are 
attributed to three factors: (1) About half of the M101 SNR candidates selected for the echelle 
observations were based on their large sizes to test whether they are superbubbles or OB/\hii\ complexes. (2) The SNRs were not visible 
in the acquisition camera, and therefore the observations were made by blind offset from nearby stars. It is possible that the 
slit was not optimally positioned to measure the expansion velocity. For example, the small SNR MF 52 may have been missed by the slit.
(3) As mentioned above, the sensitivity of the observations limits our ability to detect the faint material with the highest velocities.  

\subsection{\ha\ Fluxes of SNR Candidates}
The \emph{HST} images are flux-calibrated, and thus we can use them to measure 
the \ha\ fluxes. We measured the background-subtracted \ha\ fluxes of the 55 SNR candidates 
and report them in Table 3. The flux measurements range from $\sim$10$^{-16}$ to $\sim$10$^{-14}$ erg s$^{-1}$ cm$^{-2}$. 
Figure 5 shows a comparison between these fluxes and those measured by \citet{MF97}. In general, 
Matonick and Fesen's \ha\ flux measurements tended to be higher than ours. The discrepancies are 
probably caused by uncertainties in the exact boundaries of SNRs and background subtraction in the 
measurements based on ground-based observations. 

Figure 6 plots the \ha\ fluxes versus the sizes of the SNR candidates in M101. There appears to be a weak trend 
of increasing \ha\ flux with increasing size, but the scatter is large. The scatter may be caused by (1) ambient interstellar density -  
a SN interacting with a dense medium will be bright in \ha, and (2) the presence of massive blue stars in superbubbles 
will photoionize the ISM and produce \ha\ emission. 

\section{Underlying Stellar Populations}
The progenitors of core-collapse SNe are massive stars, which are preferentially, but not exclusively, associated with star-forming 
regions and other massive stars. The progenitors of Type Ia SNe are white dwarfs that belong to Population II. 
Therefore, SNRs' underlying stellar population has been used to diagnose the types of their SNe 
\citep{CK88b,Betal09}. For instance, a lack of nearby massive stars suggests a remnant more likely originated from a WD binary while 
the presence of massive stars suggests a remnant more likely results from a core-collapse SN. Superbubbles encompass large numbers of massive stars, such as 
OB associations or clusters, and can thus be distinguished from large bona-fide SNRs.

The \emph{HST} continuum band images show the distribution of stars surrounding the SNRs, and the 
photometric measurements of the stars can be used to assess the stellar masses and evolutionary stages. 
To evaluate the massive star population, and to minimize the foreground galactic contamination, we used the 
[F435W]$-$[F555W] color to select massive stars. We first plot evolutionary tracks for stars with a range of masses 
\citep{LS01} in the [F555W] versus [F435W]$-$[F555W] CMD for the distance of M101 (see 
Figure 7). We used [F435W]$-$[F555W] $\leq$ 0 to select stars with masses $>$ 9 $M_\odot$. For 9 $M_\odot$ stars, 
we use the [F555W] and [F435W]$-$[F555W] boundary curve that follows the evolutionary track of a 
9 $M_\odot$ star, but allow for photometric errors. These selection criteria are plotted as a dashed curve in Figure 7. 

For each SNR, the massive stars identified in the field are plotted in its third panel in Figure 2. To quantify the underlying massive 
star population, we have counted the massive stars within the boundary of each SNR candidate (Column 7 of Table 3) and 
those within 100 pc from the center of each SNR candidate, or 170 pc for remnants with radii greater than 50 pc (Column 8 of Table 3). 
The number of blue massive stars may be underestimated when a 
dense cluster is present and the photometry is incomplete. We have visually examined the broad-band images to 
identify the presence of clusters and noted them in Column 9; we have also added a '+' to the number of stars in 
Columns 7 and 8 of Table 3. 

As masssive runaway stars have been detected at distances of $\sim$100 pc from clusters \citep{Getal10}, we consider that SNRs within 100 pc 
of clusters result from core-collapse SNe. 
Several of the SNR candidates have no OB stars within a 100 pc radius, suggesting 
that they may be the result of Type Ia SNe, which typically occur in isolated areas of a galaxy where the ISM is less dense. All of the candidates 
we classify as Type Ia SNRs have a diameter less than 70 pc. In contrast, many large SNR candidates have a high density of massive stars and OB 
associations, suggesting that these large candidates may be superbubbles rather than SNRs. 

\section{X-ray and Radio Observations of SNR Candidates}
Conventionally, SNRs are confirmed by the simultaneous presence of a high \sii/\ha\ ratio, X-ray emission, 
and non-thermal radio emission \citep[e.g.,][]{C97}. Thus X-ray observations of each SNR candidate can help determine their 
nature. Known SNRs typically have X-ray luminosities ranging from 10$^{35}$ to a few $\times$ 10$^{36}$ erg s$^{-1}$ \citep{Metal83}. 
The X-rays are dominated by thermal plasma emission, except in some young core-collapse SNRs, where pulsars and pulsar wind 
nebulae may contribute a significant power-law component. The two brightest known SNRs are MF 16 in NGC 6946 and an O-rich SNR in NGC 4449, with X-ray 
luminosities of 8 $\times$ 10$^{39}$ and 2.4 $\times$ 10$^{38}$ ergs s$^{-1}$, respectively \citep{Hetal03,PF03}. While the 
X-ray spectrum of the SNR in NGC 4449 is dominated by thermal plasma emission, that of MF 16 in NGC 6946 is deficient in line 
emission and is best described by a power-law originating from a high-mass X-ray binary with a black hole primary \citep{Hetal03}. 
Superbubbles with interior SNRs interacting with their shell walls may also exhibit diffuse X-ray emission with luminosity ranges 
similar to those of conventional SNRs \citep{CM90}. 

A deep \emph{Chandra} X-ray mosaic of M101 has been made by \citet{KS10}. We have used this X-ray mosaic image 
to search for X-ray emission from the SNR candidates and found that 21 of the 55 SNR candidates in this study 
have X-ray counterparts. We have further examined their spectra and where possible classified them as thermal or power-law. These 
results are noted in Column 10 of Table 3. The X-ray fluxes of the brightest SNR candidates were previously reported by 
\citet{Petal01} and \citet{Petal07} based on a 98.2 ks \emph{Chandra} observation, and by \citet{Jetal05} based on a 42.8 ks \emph{XMM-Newton} observation. 
Using the distance of 7.4 Mpc, we have computed the X-ray luminosities 
of MF 33, MF 34, MF 46, MF 49, MF 50, MF 54, MF 65, and MF 83. Most of these X-ray counterparts have luminosities of a few $\times$ 10$^{36}$ 
ergs s$^{-1}$, consistent with those of known SNRs in the Galaxy and the MCs, with the exceptions of MF 65 and MF 83. 
MF 65 has an X-ray counterpart with a luminosity of $\sim$3$\times$10$^{37}$ ergs s$^{-1}$ (not corrected for absorption) and an ambiguous spectral shape that could be a combination of 
thermal plasma and power-law components. The X-ray counterpart to MF 83 has a luminosity of $\sim$2$\times$10$^{38}$ ergs s$^{-1}$ (not corrected for absorption), and has been identified 
to be an ultra-luminous X-ray source, instead of plasma emission \citep{Metal03}. 

Using a lower resolution \emph{ROSAT} X-ray observation of M101, \citet{W99} suggested MF 37, MF 54, MF 57, MF 83, and NGC 5471B 
as candidate hypernova remnants because of their high X-ray luminosities. We have used the deep \emph{Chandra} observation \citep{KS10} to examine 
these objects, except for NGC 5471B, which was severely off-axis in this observation. The X-ray counterpart of MF 37 has a power-law spectrum. 
As we do not have a good \emph{HST} \ha\ image showing the SNR, we cannot assess the nature of MF 37. As pointed out by \citet{Setal01}, MF 54 is coincident with a 
faint X-ray source 6-7\arcsec\ from a very bright X-ray source which was misidentified as the X-ray counterpart to MF 54 by \citet{W99}. 
Likewise, MF 57 also had its X-ray counterpart misidentified in the lower resolution \emph{ROSAT} observation \citep{Setal01}. Using the better coordinates 
of MF 57 determined from the high-resolution \emph{HST} \ha\ image, we were able to identify a fainter X-ray counterpart of MF 57, but with too few counts for 
a spectral fit. MF 83 is a superbubble powered by an ultra-luminous X-ray source, 
as discussed above. None of these four SNR candidates qualify as hypernova remnants. 

None of the SNR candidates optically identified by \citet{MF97} had radio counterparts in the VLA 20-cm observations by \citet{Eetal02}. This 
is not surprising, because the detection limit of this radio observation corresponds to a young SNR such as Cas A. 
To search for small, young SNRs that have been missed in the optical survey, we have compared radio and X-ray observations and found 12 radio sources 
coincident with X-ray counterparts. Seven of these sources have \emph{HST} \ha\ and broad-band images for more detailed examination: M101-E-$\alpha$, M101-E-$\gamma$, 
M101-E-$\epsilon$, M101-E-$\theta$, M101-E-$\phi$, M101-S-$\alpha$, and M101-S-$\gamma$ (designations from \citet{Eetal02}). We find that the radio sources M101-E-$\alpha$ 
and M101-E-$\gamma$ correspond to bright nebular knots in the giant \hii\ 
region NGC 5462. M101-E-$\epsilon$ and M101-E-$\theta$ are on the outskirts of the giant \hii\ region NGC 5462; the former has no distinct 
\ha\ emission that can be identified as a SNR shell, while the latter has a resolved optical counterpart that is dominated by continuum emission, and thus is likely a background galaxy. 
M101-E-$\phi$ corresponds to a bright nebular knot in the giant \hii\ region NGC 5461, similar to M101-E-$\alpha$ and M101-E-$\gamma$. 
M101-S-$\alpha$ corresponds to SN1970G, the known radio SN in the giant \hii\ region NGC 5455; the X-ray emission from SN1970G is unambiguously detected \citep{IK05}. 
M101-S-$\gamma$ has a small offset between the X-ray and radio sources, and the \emph{HST} continuum images show that the X-ray source is coincident with a background 
spiral galaxy and the radio source is coincident with a background elliptical galaxy. None of these seven objects can be identified as Cas A type SNRs. 

We also examined the X-ray data for the 38 SNR candidates identified by \citet{MF97} 
that do not have archived optical \emph{HST} images available. It should be noted that since \emph{HST} 
images were unavailable, we were unable to measure accurate coordinates and sizes for these SNR candidates, so 
we resorted to using the coordinates and sizes published by \citet{MF97} for each 
SNR candidate. Since their observations were ground-based and thus had poorer resolution 
than the \emph{HST} images, many of the positions and sizes of the SNR candidates may not be 
very accurate. Thus there is an additional degree of uncertainty in the coincidence between X-ray 
sources and the SNR candidates. We find probable coincident X-ray sources for 11 objects: MF 20, MF 21, MF 22, 
MF 24, MF 29 MF 37, MF 61, MF 64, MF 71, MF 77, and MF 93. The other 27 objects do not appear to have X-ray counterparts. 
Without high-resoultion \emph{HST} \ha\ images of these 38 candidates, we cannot further assess 
the nature of these SNR candidates. 

\section{Nature of SNR Candidates}
Based on the morphology revealed in optical images, kinematic information derived from high-dispersion 
echelle spectra, stellar population information extracted from the photometric data, and the shocked hot gas 
shown in X-ray data, we attempt to distinguish among Type Ia SNe, core-collapse SNe, and superbubbles 
produced by OB associations or large numbers of massive stars. Our proposed classifications 
are given in Column 12 of Table 3, while a more detailed explanation of the justifications for each classification 
is given in Appendix A. As our classification is based on statistical analysis, we cannot confirm our assessment 
of the nature of each SNR candidate, so we have assigned a confidence value A, B, or C, roughly corresponding to 
``most likely,'' ``probable,'' and ``possible,'' respectively (Column 13 of Table 3).

\subsection{Superbubbles}
A significant fraction of the SNR candidates in M101 have sizes $\geq$ 100 pc, much larger than 
typical SNRs in the Galaxy and the MCs. This is surprising because a single SNR with a diameter 
$\geq$ 100 pc would likely be located in a low-density medium, resulting in a low surface brightness that would be difficult 
to detect in ground-based optical images. It is possible that some of these large SNR candidates are superbubbles 
blown by OB associations or clusters. Superbubbles can be distinguished from SNRs by the presence 
of a visible blue cluster or a large number of OB stars within the boundary of the SNR candidate, and a 
slow \ha\ expansion velocity ($\leq$ 100 \kms). Therefore, we first searched for OB associations or clusters 
within the SNR boundaries and identified the associated large shells as superbubbles. 
To assess whether a superbubble contains a SNR in its interior, we examined the \ha\ expansion velocity 
and X-ray observations. If a superbubble shows high-velocity shocked material or diffuse X-ray emission, it is likely 
that the superbubble contains the remnant of a recent SN explosion. Only one of the large SNR candidates, MF 83, 
has both an internal cluster and exhibits bright X-ray emission and was suggested to be a hypernova remnant 
\citep{W99}, but the X-ray emission originates from an unresolved ultra-luminous X-ray source \citep{Metal03}, 
which might be responsible for the moderately high expansion velocity of the shell, $\sim$50 \kms\ \citep{Letal01}.
For large SNR candidates without visible clusters, if we find the expansion velocity to be lower than 100 \kms, we also 
classify the object as a superbubble.

Young superbubbles may be smaller than 100 pc; therefore, we have examined the underlying stellar populations 
of all of the SNR candidates, and indeed found some objects smaller than 100 pc that also have clusters or 
OB associations present. It is possible that the optical shell structure represents a superbubble blown by the OB 
association. If there has been a recent SN explosion near the superbubble shell wall, the interaction between the SNR 
shock and the shell wall will produce bright X-ray emission \citep{CM90}. Thus, we also examined the X-ray 
observations of these objects to confirm the presence of a SNR, and used the echelle data, when available, to search 
for high-velocity shocked gas. Some SNR candidates are small superbubbles with interior SN explosions and 
it is a matter of semantics to call these objects SNRs or superbubbles. For these SNR candidates we have classified them 
as SNRs in Table 3 but their superbubbles are noted in the individual descriptions in the Appendix. 

\subsection{SNRs of Core-Collapse SNe versus Type Ia SNe}
For the SNR candidates not identified as superbubbles, we attempted to distinguish between SNRs resulting from Type Ia 
SNe and core-collapse SNe. The criteria we used for identifying SNRs resulting from core-collapse SNe were: 
(1) the presence of many OB stars or even a cluster in or around the SNR candidate, and (2) a non-uniform shell, often 
brighter on one side and often having visible \ha\ emission outside of the boundary of the SNR candidate (from gas 
photoionized by neaby OB stars). These two criteria reflect an environment rich in massive stars and photoionized gas, 
and SNe in these environments are more likely to have massive star progenitors. 
We can further estimate the uncertainty of this classification from the statistics of Galactic O stars. 
Among the 227 Galactic O stars with $V <$ 8,  $\sim$83\% are in clusters, $\sim$10\% are runaways, 
and only 5--10\% are truly isolated \citep{Detal04,Detal05,ZY07}. 
Therefore, about 10$\pm$5\% of core-collapse SNRs are not associated with other massive stars or star-forming regions, and 
will not be identified as such by our criteria.

Identifying characteristics of SNRs originating from Type Ia SNe include: (1) the absence of nearby clusters or OB stars, 
(2) a round, uniform shell morphology, and (3) a relatively low \ha\ flux, as compared to SNRs originating from core-collapse 
SNe. These criteria are based on our knowledge of the Type Ia SNRs in the Large Magellanic Cloud 
(LMC). Most of the Type Ia SNRs in the LMC are more or less round, without large brightness variations along the rim 
\citep{Betal06,Metal83}. As Type Ia SNe typically occur in a low-density medium, the surface brightness is usually 
lower. However, there are exceptions, for example, N103B in the LMC is located near a \hii\ region in a crowded stellar 
environment, yet the abundances determined from the X-ray spectra reveal that it is the result of a Type Ia SN \citep{Hetal95}.
As there are 7 confirmed and 6 possible Type Ia SNRs in the LMC \citep{CK88b,Hetal95,Betal06}, and N103B is the only one that is 
interacting with \hii\ regions, we suggest that our criteria have a misidentification rate of $\sim$7\%. This is consistent 
with the 5--8\% of Type Ia SNe whose spectra show nebular \ha\ emission, indicating the existence of an \hii\ region (J.\ Silverman 2011, personal communication).

We do not compare the Type Ia SNRs in M101 with those in the Galaxy because of the uncertain distances and 
the large, non-uniform foreground extinction in the Galactic plane. The Type Ia SNRs DEM L238 and DEM L249 in the LMC, on 
the other hand, have sizes of 39$\times$33 pc and 45$\times$30 pc, respectively, and have a small foreground extinction; 
thus these SNRs can be compared with those we classify as having resulted from Type Ia SNe in M101. The \ha\ luminosities of DEM L238 and DEM L249 
are 1.8$\times$10$^{36}$ and 2.4$\times$10$^{36}$ ergs s$^{-1}$, respectively \citep{KH86}. Using the digital data from 
\citet{KH86}, we find the peak emission measures of these two SNRs to be $\sim$900 cm$^{-6}$ pc. For comparison, the M101 Type Ia SNR candidate 
MF 23 has an \ha\ luminosity of 1.4$\times$10$^{37}$ ergs s$^{-1}$, a diameter of 70 pc, and a peak emission measure of $\sim$700 cm$^{-6}$ pc. 
The detectibility of a nebula depends on its surface brightness or the peak emission measure. DEM L238 and DEM L249, having 
higher surface brightnesses than MF 23, can easily be detected in the \emph{HST} \ha\ images used in our study. 

In M101, the SNR MF 33 may be similar in nature to the Type Ia SNR N103B. Both are located near a very densely populated 
stellar neighborhood, exhibit very bright X-ray emission, and have a shell that is brighter on one side. However, MF 33 also 
resembles the ultra-luminous core-collapse SNR MF 16 in NGC 6946 \citep{DGC00}. It is difficult to determine the real nature of 
MF 33 unless an X-ray or optical spectrum adequate for spectral analysis to determine abundances can be obtained. We have listed 
the classification of this object as uncertain, since we were not able to determine whether it resulted from a core-collapse SN 
or Type Ia SN.

\subsection{OB/\hii\ Complexes}
Several of the SNR candidates did not have a clearly defined shell structure in the \emph{HST} 
\ha\ images. The absence of a defined shell suggests that these are neither SNRs nor superbubbles, 
but rather OB/\hii\ complexes with a localized enhancement in \sii\ emission, which allowed them to be detected in 
Matonick and Fesen's optical survey, since their criteria for identifying nebulae as SNR 
candidates was a \sii/\ha\ $\geq$ 0.45. These SNR candidates may be a combination of multiple objects, but 
since \emph{HST} \sii\ images are not available, we cannot identify exactly what part of the 
region was detected to have a high \sii/\ha\ ratio in the \citet{MF97} survey, and the ground-based images 
they published do not have high enough resolution to make this distinction. It is possible that there 
is a SNR located somewhere within the OB/\hii\ complexes, but since we do not see a shell structure 
in the \ha\ images, we cannot confirm its presence. Four of the nine OB/\hii\ complexes we identified are 
located on the outskirts of the giant \hii\ region NGC 5462 on the east edge of M101. The \sii\ enhancements 
in these regions are similar to those commonly seen outside of active star forming regions in irregular 
galaxies \citep{HG90}.

Based on studies by \citet{H94} and \citet{WB94}, \citet{MF97} suggested that objects having a 
\sii/\ha\ $\leq$ 0.60 and a diameter $\geq$ 100 pc may not be SNRs. They classified 10 objects in 
M101 fitting those criteria, 5 of which we included in our study, MF 01, MF 12, MF 51, MF 67, and MF 78. 
We have identified MF 01, MF 51, and MF 78 as superbubbles, and MF 67 as an OB/\hii\ complex, supporting 
the hypothesis made by Matonick and Fesen. However, we believe that MF 12 is the result of a core-collapse SN. 
MF 12 is located near a substantial \hii\ region, which likely made identifying the 
correct size of the remnant in ground-based images difficult. In our \emph{HST} images, we are able to 
differentiate between the \hii\ region and the SNR candidate, and have measured MF 12 to have a size 
of $\sim$60 pc, placing it below the 100 pc threshold.

\subsection{Distribution of SNR Types}
We suggest that $\sim$60\% (34 of 55) of the SNR candidates we studied are true, conventional SNRs. This indicates 
that although the \sii/\ha\ ratio diagnostic of SNRs remains the most sensitive way to identify extragalactic 
SNRs, an examination of the physical properties and stellar environments of SNR candidates is necessary 
to eliminate contamination from non-SNR objects that may have a high \sii/\ha\ ratio. Superbubbles are the 
most prevalent of the non-SNR objects we identified from the Matonick and Fesen list, making up 
$\sim$20\% (11 of 55) of the SNR candidates we studied. The remaining 10 objects were somewhat ambiguous. We identified 
most of them as OB/\hii\ complexes with no clearly defined shell. MF 85 was classified as diffuse \hii\ emission, as there 
does not appear to be a visible shell structure or OB association. 

We estimate that $\sim$25\% (9 of 34) of the SNRs we identified are the result of Type Ia SNe. 
This can be compared with the Type Ia SNRs in the LMC. Using a similar stellar population identification criterion, 12 of the 45 ($\sim$27\%) known SNRs in 
the LMC have been suggested to result from Type Ia SNe \citep{CK88b,Detal10}. 
There are large differences in the galactic types (with or without spiral arms) and observational biases.
For example, most of the known Type Ia SNRs in the LMC have a diameter less than 30 pc, 
whereas those in M101 are larger because the smallest SNRs ($<$ 20 pc) in M101 have not yet been identified.
Despite all these differences, the percentages of Type Ia SNRs show close agreement between the LMC and M101. 
Note that the fractions of Type Ia SNRs of the LMC and M101 are much higher than the fraction of Type Ia SNe 
observed in late-type galaxies (Scd spirals and irregulars) because a significant fraction of core-collapse SNe occur is superbubbles and do not 
necessarily develop recognizable SNRs.

\subsection{Spatial Distribution of SNRs in M101}
To study the distribution of SNR candidates in M101, 
we visually inspected the images to assess whether each candidate was in or between spiral arms (See Figure 1 and Column 11 of Table 3). 
Thirty-one of the SNR candidates are located in the spiral arms of M101, while 21 are located in 
interarm regions, giving a ratio of $\sim$1.5:1. The presence of dense interstellar gas and a high stellar 
density in the spiral arms of M101 would seem to favor the occurences of SNRs in arm regions. However, the presence 
of clusters and \hii\ regions in the spiral arms may obscure many SNRs, making detection difficult. None of the 
objects we classify as bona-fide SNRs are located inside of \hii\ regions, and only one, MF 12, lies on the outskirts of an \hii\ region. 
The SNR candidates located in interarm regions are much easier to detect and thus make up a larger portion of 
the SNR sample in M101 than expected. Matonick and Fesen hypothesized that the number of SNRs in M101 may be 
up to four times larger than the 93 they detected.  

We have also analyzed the types of SNRs found in both the arm and interarm regions. Of the 11 SNRs we suggest 
to be the result of Type Ia SNe, only two are found in the spiral arms of M101. Conversely, 55\%\ (14 of 25) 
of the SNRs we suggest to be the result of core-collapse SNe are located in the spiral arms, suggesting the 
presence of more high-mass stars in the spiral arms of M101. Superbubbles also tend to be located in the 
spiral arms; only four of the 11 we identified are located in interarm regions. 

\section{Summary}
We have examined \emph{HST} \ha\ and broad-band images and photometry, high-dispersion 
echelle spectra, and \emph{Chandra} X-ray data to study the 
nature and distribution of SNRs in M101. Of the 93 SNR candidates indentifed by \citet{MF97}, 55 had both \emph{HST} 
and \emph{Chandra} observations available. For these objects, we measured positions and sizes using the \emph{HST} \ha\ image 
and used \emph{HST} broad-band images and photometry to examine their underlying interstellar environment and stellar population. The 
\emph{Chandra} data were then used to search for X-ray counterparts to the SNR candidates. Finally, high-dispersion echelle spectra were 
used to measure the expansion velocities for 18 of the SNR candidates. 

We then used this information to assess the nature of each SNR candidate. 
Large objects with a large number of nearby massive stars or clusters within the boundary of the SNR candidate also typically 
had low expansion velocities ($\leq$ 100 \kms). These were classified as superbubbles. Smaller objects with higher expansion velocities ($\geq$ 100 \kms) 
and thermal X-ray emission were typically classified as either core-collapse or Type Ia SNRs. In such cases, the stellar and interstellar environment played 
a critical role in our final diagnosis. Objects with a large concentration 
of nearby massive stars were typically classfied as remnants of core-collapse SNe, while objects 
in isolated areas of M101 with no or few nearby massive stars were classified as remnants of Type Ia SNe. Finally, objects encompassing 
massive stars and ionized gas, but without a 
clear shell structure were classified as OB/\hii\ complexes, which may harbor a SNR, 
but without further observations, their nature cannot be assessed. 

Of the 55 SNR candidates we analyzed, nine are likely remnants of Type Ia SNe, 
25 remnants of core-collapse SNe, 11 superbubbles, eight OB/\hii\ complexes, and 
one diffuse \hii\ region. We were uncertain as to the nature of MF 33, which may be argued to result from either a 
core-collapse or Type Ia SN. The remaining 38 objects in the \citet{MF97} catalog did not have the \emph{HST} \ha\ images neccessary to assess their properties and nature. 

We have compared these 55 SNR candidates to a list of published radio sources in M101, but the radio observations are not deep enough to detect these SNR candidates. 
To search for possible SNR candidates missed by the initial optical survey, we have also compared the radio sources to X-ray point sources in M101. Seven 
radio sources are coincident with X-ray point sources, however none of them correspond to potential SNRs. Currently, the X-ray and radio observations available 
lack the sensitivity to be used to confirm most SNRs identified with \sii\ and \ha\ images. 

Our work has demonstrated that \emph{HST} \ha\ and continuum images are essential in revealing the physical structure and underlying stellar 
population of SNR candidates in M101 to enable further investigation of their nature. The archival \emph{HST} \ha\ images do not cover the entire 
galaxy, therefore a global study of M101 is currently impossible. Furthermore, \emph{HST} \sii\ images are needed for comparison with \ha\ images 
to search for new small SNR candidates which could not be identified by ground-based observations. Only after a complete census of SNRs is obtained 
can we attempt a comprehensive investigation of the distribution, population, and rates of SNe in this galaxy. 

\acknowledgments 
This research is supported in part by NASA grants NNX11AH96G and SAO GO0-11025X. We thank the referee, 
Dr. Eric Schlegel for constructive suggestions to improve the paper. 

This paper has used observations made with the NASA/ESA \emph{Hubble Space Telescope}, and obtained from the data 
archive at the Space Telescope Science Institute and the Hubble Legacy Archive, which is a 
collaboration between the Space Telescope Science Institute (STScI/NASA), the Space Telescope 
European Coordinating Facility (ST-ECF/ESA) and the Canadian Astronomy Data Centre (CADC/NRC/CSA). 
STScI is operated by the Association of Universities for Research in Astronomy, Inc. under NASA 
contract NAS 5-26555.

\clearpage

\appendix
\section{Summary of M101 SNR Classification}
For each SNR candidate, the object name, classification, confidence level, and remarks are given below.

\indent\emph{MF 01:} Superbubble - A - MF 01 is very large, with a diameter $\geq$ 100 pc, and a visible cluster located inside remnant. It also has a high \ha\ flux. \\
\indent\emph{MF 07:} Superbubble - B - MF 07 encompasses a sparse group of massive stars. The shell is likely a superbubble blown jointly by SNe and stellar winds from this group of massive stars. \\
\indent\emph{MF 08:} Type Ia - C - MF 08 has one massive star projected within its boundary, consistent with the background population. Its uniform circular morphology is similar to Type Ia SNRs in the Galaxy or the MCs, however it has a large diameter. \\
\indent\emph{MF 09:} Core-Collapse - A - MF 09 is in an environment with a high concentration of massive stars. The shell is non-uniform, indicating interaction with an inhomogeneous ambient ISM. \\
\indent\emph{MF 11:} Core-Collapse - A - MF 11 is associated with a \hii\ region and small cluster of stars that are not well resolved in the \emph{HST} images for photometry, suggesting a core-collapse SN. \\
\indent\emph{MF 12:} Core-Collapse - A - MF 12 is has very bright \ha\ emission, is located near an \hii\ region, encompasses a small cluster of stars, and has bright X-ray emission. \\
\indent\emph{MF 13:} Core-Collapse - A - MF 13 has a non-uniform shell. It encompasses a small concentration of stars, and has visible X-ray emission. \\
\indent\emph{MF 15:} Core-Collapse - B - MF 15 has an irregular morphology, making its identification as a SNR uncertain. It is near several massive stars, therefore if it is indeed a SNR, its progenitor is likely a core-collapse SN. \\
\indent\emph{MF 16:} Core-Collapse - B - MF 16 has massive stars located within the boundary of the remnant, and its morphology is non-uniform, indicating it is likely the result of a core-collapse SN. \\
\indent\emph{MF 17:} Core-Collapse - A - MF 17 has a cluster of massive stars inside the remnant's boundary. Its shell is non-uniform. We cannot exclude the possibility that it is a superbubble with an interior SN explosion. Either way, a core-collapse SN has occured in MF 17. \\
\indent\emph{MF 19:} Superbubble - A - MF 19 is a very large shell with a visible cluster of massive stars within its boundary, suggesting it is a superbubble. \\
\indent\emph{MF 23:} Type Ia - B - MF 23 is not associated with any massive stars or \hii\ regions. \\
\indent\emph{MF 25:} Core-Collapse - A - MF 25 has a non-uniform shell, is associated with \hii\ regions, and has a high expansion velocity. A cluster within MF 25 can be clearly seen in the F606W image, though other broad band images are not available to study the population of massive stars. \\
\indent\emph{MF 26:} Type Ia - A - MF 26 has a very small, uniform shell. It is located in a region with a very low concentration of massive stars, making it a good candidate for a Type Ia SN. \\
\indent\emph{MF 27:} Core-Collapse - A - MF 27's shell is non-uniform, and it encompasses a few stars. Faint X-ray emission is also present. These factors suggest a high mass progenitor. \\
\indent\emph{MF 28:} Core-Collapse - C - MF 28 has an irregular morphology. A small concentration of stars within the boundary is present in the F606W image, however we do not have other broad band images to study the underlying stellar population in greater detail. If it is indeed a SNR, it is most likely the remnant of a core-collapse SN. \\
\indent\emph{MF 30:} Type Ia - A - MF 30 has a small, round shell. It encompasses a few stars, though there are no nearby massive stars, and it is not associated with an \hii\ region. \\
\indent\emph{MF 32:} Core-Collapse - C - This remnant has a clover leaf shaped morphology and bright X-ray emission. It is associated with diffuse \hii\ regions superposed on a dust lane along the inner edge of a spiral arm. While we classify it as a remnant of a core-collapse SN, it is possible that MF 32 is a M101 analog of the Type Ia SNR N103B in the LMC \citep{CK88b}. N103B is near an \hii\ region, and superposed on a molecular cloud, but its X-ray spectrum shows abundances consistent with a Type Ia SN explosion \citep{Hetal95}. \\
\indent\emph{MF 33:} Uncertain - D - MF 33 is small with a non-uniform shell, and exhibits bright \ha\ and X-ray emission. It is located near \hii\ regions, but does not encompass any massive stars. MF 33 resembles both the Type Ia SNR N103B in the LMC \citep{Hetal95} and the core-collapse SNR MF 16 in NGC 6946 \citep{DGC00}, so we cannot classify it as either Type Ia or core-collapse. \\
\indent\emph{MF 34:} Core-Collapse - A - MF 34 has an triangular morphology, and it is located near several massive stars and \hii\ regions. It has visible X-ray emission. \\
\indent\emph{MF 40:} Core-Collapse - A - MF 40 has a non-uniform shell and diffuse \ha\ emission surrounding the SNR. It encloses a small concentration of stars, including a massive star. It is associated with \hii\ regions and has visible X-ray emission. \\
\indent\emph{MF 41:} Core-Collapse - A - MF 41 is located in an environment rich in massive stars. It has a triangular morphology, and is in a diffuse \ha\ emission field on the outskirts of a large \hii\ complex. \\
\indent\emph{MF 42:} Type Ia - A - MF 42 is in an environment devoid of massive stars. It is not associated with any diffuse \ha\ emission. \\
\indent\emph{MF 43:} Type Ia - A - MF 43 has a double-lobed morphology, reminiscent of the SNR DEM L316 in the LMC \citep{WC05}. It is located in an environment devoid of massive stars and ionized gas. This remnant's progenitor is most likely a Type Ia SN, similar to that of DEM L316a. \\
\indent\emph{MF 44:} Core-Collapse - A - MF 44 has a bipolar structure and surrounding diffuse \ha\ emission. It encompasses several stars, is located in a field rich in massive stars, and has visible X-ray emission. \\
\indent\emph{MF 45:} OB/\hii\ Complex - A - MF 45 has no clearly defined shell structure. It is diffuse with a few bright knots. It is located in an environment rich in massive stars and ionized gas. This object is most likely a OB/\hii\ complex that contains a core-collapse SN explosion. \\
\indent\emph{MF 46:} Type Ia - A - MF 46 has a round, moderately uniform shell and is located in a field with no massive stars. \\
\indent\emph{MF 47:} Superbubble - A - This large shell has a high concentration of stars, some of which are massive, within its boundary, and is located near another concentration of stars and ionized gas. \\
\indent\emph{MF 48:} OB/\hii\ Complex - A - MF 48 does not have a defined shell structure. It has bright knots superimposed on diffuse emission. The environment is rich in ionized gas and massive stars, indicating that this object is an OB/\hii\ complex that may contain a core-collapse SN. \\
\indent\emph{MF 49:} Core-Collapse - A - MF 49 has a very high surface brightness, and bright X-ray emission. A Type Ia SNR with the size of MF 49, 19 pc, would still be young enough to be Balmer line dominant with little forbidden line emission, such as DEM L71 in the LMC \citep{Metal83}. As MF 49 was identified from its high \sii/\ha\ ratio, it must be a core-collapse SNR. \\
\indent\emph{MF 50:} Core-Collapse - A - MF 50 is similar to MF 49 in every respect and thus is also classified as a core-collapse SNR. \\
\indent\emph{MF 51:} Superbubble - A - MF 51 has a diameter $\geq$ 100 pc, and encompasses several massive stars. These are characteristics of a superbubble. \\
\indent\emph{MF 52:} Core-Collapse - A - MF 52 has an irregular morphology, and lacks bright X-ray emission. There are some massive stars located in the vicinity of the remnant. \\
\indent\emph{MF 53:} Superbubble - A - MF 53 has a large number of massive stars enclosed within its shell. This, along with the very large size, suggest that MF 53 is a superbubble. \\
\indent\emph{MF 54:} Core-Collapse - A - MF 54 has a small cluster of massive stars within its boundary, and is surrounded by diffuse ionized gas. Its shell is not uniformly shaped. \\
\indent\emph{MF 57:} Type Ia - A - MF 54 has a small, fairly uniform shell, and is located in an environment with very few stars. It also exhibits faint X-ray emission. This is consistent with remnants resulting from Type Ia SNe in the MCs. \\
\indent\emph{MF 58:} Superbubble - A - MF 58 is very large and of somewhat irregular shape. It is located among a collection of \hii\ regions. This large size suggests MF 58 is an old superbubble whose most massive main-sequence stars are late B stars. \\
\indent\emph{MF 59:} Core-Collapse - A - MF 59 is located near \hii\ regions, and its shell brightness is not uniform. There are some nearby massive stars, and a lack of bright X-ray emission. \\
\indent\emph{MF 60:} Core-Collapse - B - MF 60's shell is small with non-uniform brightness. It is located near some massive stars and diffuse \ha\ emission, but lacks bright X-ray emission. \\
\indent\emph{MF 62:} Core-Collapse - B - MF 62 is located in a dense stellar environment, with massive stars in the surroundings. This object is likely a remnant of a core-collapse SN, or a superbubble with an interior core-collapse SN explosion. \\
\indent\emph{MF 65:} Core-Collapse - A - MF 65 encompasses a small cluster, and is surrounded by diffuse ionized gas. It is coincident with a bright X-ray source, but it is ambiguous whether the source is thermal or non-thermal. The X-ray source is coincident with arcs at 1\arcsec\ from the \ha\ peak. \\
\indent\emph{MF 67:} OB/\hii\ Complex - C - Within the boundary of MF 67, there are multiple shell structures and regions of diffuse emission. It is not clear whether there is an SNR and which structure corresponds to the SNR. If there is a SNR within the boundary of MF 67, it most likely resulted from a core-collapse SN. \\
\indent\emph{MF 69:} OB/\hii\ Complex - C - MF 69 is a large diffuse emission region with no defined shell structure. This morphology and the distribution of massive stars in the vicinity suggests that this is an OB/\hii\ complex. \\
\indent\emph{MF 73:} Superbubble - A - The large shell size and distribution of massive stars suggests this object is a superbubble. \\
\indent\emph{MF 75:} Superbubble - A - The large number of massive stars within this large shell suggests that MF 75 is a superbubble. \\
\indent\emph{MF 76:} Core-Collapse - A - MF 76 is located near the superbubble MF 75, and has neighboring massive stars. \\
\indent\emph{MF 78:} OB/\hii\ Complex - C - This object is diffuse without a defined shell structure. It encompasses a large number of massive stars. It is not clear whether there is a SNR at all. \\
\indent\emph{MF 81:} Type Ia - A - MF 81 has a small, uniform shell, and is located in an environment with very few stars. It exhibits X-ray emission. \\
\indent\emph{MF 82:} Superbubble - C - It is not clear whether MF 82 is a superbubble or an OB/\hii\ complex since there are no continuum images available. The interstellar environment suggests the existance of OB stars in its vicinity. \\
\indent\emph{MF 83:} Superbubble - A - MF 83 is a superbubble around an ultra-luminous X-ray source \citep{Letal01,Metal03}. \\
\indent\emph{MF 85:} Diffuse \hii\ - C - This object shows only diffuse emission without any shell structure. There are massive stars within and in the vicinity of MF 85. It is not clear whether there is a SNR present. \\
\indent\emph{MF 88:} Core-Collapse - A - MF 88 is located near a large number of massive stars and has a well-defined shell structure. \\
\indent\emph{MF 89:} OB/\hii\ Complex - A - This object is at the southern tip of the giant \hii\ region NGC 5462. It has a very high number of associated OB stars, and lacks a clearly defined shell. \\
\indent\emph{MF 90:} OB/\hii\ Complex - A - This object is also at the southern tip of the giant \hii\ region NGC 5462. It has several nearby OB stars, and no defined shell. \\
\indent\emph{MF 91:} OB/\hii\ Complex - C - This object is on the southeast outskirts of NGC 5462. This object neighbors several massive stars. It may contain a SNR, but there is no identifiable shell structure. \\

\clearpage

\begin{deluxetable}{cccc}
\tabletypesize{\scriptsize}
\tablewidth{0pc}
\tablecolumns{4}
\tablecaption{List of \emph{HST} Archived Images}
\tablehead{
\colhead{Proposal ID} & \colhead{PI Name} & \colhead{Instrument} & \colhead{Bands}}
\startdata
5210 & Trauger & WFPC2 & F656N \\
6044 & Garnett & WFPC2 & F656N \\
6713 & Sparks & WFPC2 & F656N, F606W \\
6829 & Chu & WFPC2 & F656N \\
8591 & Richstone & WFPC2 & F656N \\
9490 & Kuntz & ACS/WFC & F435W, F555W, F814W \\
9492 & Bresolin & ACS/WFC & F435W, F555W, F658N, F814W \\
9720 & Chandar & ACS/WFC & F658N \\
10918 & Freedman & ACS/WFC & F555W, F814W \\
\enddata
\end{deluxetable}

\clearpage

\begin{deluxetable}{ccccc}
\tabletypesize{\scriptsize}
\tablewidth{0pc}
\tablecolumns{5}
\tablecaption{Journal of KPNO 4 m Echelle Observations}
\tablehead{
\colhead{MF} & \colhead{Date of} & \colhead{Slit} & \colhead{Exposure} & \colhead{Instrumental} \\
\colhead{No.} & \colhead{Observation} & \colhead{Width} & \colhead{Time} & \colhead{FWHM}}
\startdata
19 & 2000 Apr 22 & 2\farcs0 & 2 $\times$ 1200 s & 16.2 \kms \\
24 & 2000 Apr 24 & 2\farcs0 & 2 $\times$ 1200 s & 16.4 \kms \\
25 & 2000 Apr 24 & 2\farcs0 & 2 $\times$ 1200 s & 15.6 \kms \\
27 & 2000 Apr 24 & 2\farcs0 & 3 $\times$ 1200 s & 16.8 \kms \\
28 & 2000 Apr 24 & 2\farcs0 & 2 $\times$ 1200 s & 16.2 \kms \\
32 & 2000 Apr 24 & 2\farcs0 & 3 $\times$ 1200 s & 16.8 \kms \\
45 & 2000 Apr 23 & 1\farcs5 & 2 $\times$ 1200 s & 13.4 \kms \\
48 & 2000 Apr 22 & 2\farcs0 & 3 $\times$ 1200 s & 16.2 \kms \\
51 & 2000 Apr 23 & 1\farcs5 & 2 $\times$ 1200 s & 11.8 \kms \\
52 & 2000 Apr 24 & 2\farcs0 & 2 $\times$ 1200 s & 15.6 \kms \\
53 & 1999 Jul 02 & 1\farcs5 & 2 $\times$ 1800 s & 14.6 \kms \\
59 & 2000 Apr 24 & 2\farcs0 & 2 $\times$ 1200 s & 15.3 \kms \\
75 & 2000 Apr 23 & 1\farcs5 & 2 $\times$ 1200 s & 11.6 \kms \\
78 & 2000 Apr 23 & 1\farcs5 & 2 $\times$ 1200 s & 13.9 \kms \\
85 & 2000 Apr 22 & 2\farcs0 & 2 $\times$ 1200 s & 16.0 \kms \\
89 & 2000 Apr 23 & 1\farcs5 & 2 $\times$ 1200 s & 13.4 \kms \\
90 & 2000 Apr 22 & 2\farcs0 & 2 $\times$ 1200 s & 15.5 \kms \\
91 & 2000 Apr 23 & 1\farcs5 & 2 $\times$ 1200 s & 12.8 \kms \\
\enddata
\end{deluxetable}

\clearpage

\begin{deluxetable}{ccccccccccccc}
\tabletypesize{\scriptsize}
\rotate
\tablewidth{0pt}
\tablecolumns{13}
\tablecaption{Properties of Candidate SNRs in M101}
\tablehead{
\colhead{MF} & \colhead{R.A.} & \colhead{Dec.} & \colhead{Diameter} & \colhead{Diameter} & \colhead{\ha\ Flux} & \colhead{Stars} & \colhead{Stars} 
& \colhead{Cluster} & \colhead{X-ray\tablenotemark{c}} & \colhead{Region} & \colhead{Proposed} & \colhead{Confidence\tablenotemark{d}}\\
\colhead{Number} & \colhead{(J2000)} & \colhead{(J2000)} & \colhead{(arcsec)} & \colhead{(pc)} & \colhead{(erg\,cm$^{-2}$\,s$^{-1}$)} & \colhead{Within} & \colhead{in the} 
& \colhead{} & \colhead{} & \colhead{} & \colhead{Classification} & \colhead{}\\
\colhead{} & \colhead{} & \colhead{} & \colhead{} & \colhead{} & \colhead{} & \colhead{SNR\tablenotemark{a}} & \colhead{Field\tablenotemark{b}} 
& \colhead{} & \colhead{} & \colhead{} & \colhead{} & \colhead{}\\
\colhead{(1)} & \colhead{(2)} & \colhead{(3)} & \colhead{(4)} & \colhead{(5)} & \colhead{(6)} & \colhead{(7)} & \colhead{(8)} 
& \colhead{(9)} & \colhead{(10)} & \colhead{(11)} & \colhead{(12)} & \colhead{(13)} 
}
\startdata
01 & 14:02:20.28 & +54:21:35.45 & 4.1 & 150 & 2.7E$-$14 & 10+ & 15+ & Yes & No & Arm & Superbubble & A \\
07 & 14:02:32.82 & +54:25:54.82 & 2.7 & 100 & 1.1E$-$15 & 5 & 5 & No & No & Interarm & Superbubble & B \\
08 & 14:02:36.42 & +54:23:57.16 & 2.1 & 80 & 1.0E$-$15 & 1 & 2 & No & No & Interarm & Type Ia & C \\
09 & 14:02:38.41 & +54:22:25.20 & 0.7 & 30 & 1.0E$-$15 & 0 & 19 & No & No & Arm & Core-Collapse & A \\
11 & 14:02:41.86 & +54:22:36.78 & 1.0 & 40 & 1.1E$-$15 & 0+ & 0+ & Yes & No & Interarm & Core-Collapse & A \\
12 & 14:02:43.86 & +54:20:05.70 & 1.0 & 40 & 1.6E$-$14 & 0+ & 11+ & Yes & Th & H\,{\sc II} & Core-Collapse & A \\
13 & 14:02:44.18 & +54:20:34.66 & 0.9 & 30 & 1.4E$-$15 & 0 & 3 & No & Am & Interarm & Core-Collapse & A \\
15 & 14:02:45.77 & +54:25:19.92 & 1.1 & 40 & 6.3E$-$16 & 1 & 6 & No & No & Interarm & Core-Collapse & B \\
16 & 14:02:46.07 & +54:24:40.83 & 1.8 & 60 & 9.2E$-$16 & 2 & 5 & No & No & Interarm & Core-Collapse & B \\
17 & 14:02:49.11 & +54:20:56.22 & 1.6 & 60 & 2.1E$-$15 & 2+ & 2+ & Yes & Th & Arm & Core-Collapse & A \\
19 & 14:02:49.76 & +54:22:46.39 & 7.1 & 250 & 6.1E$-$15 & 6+ & 7+ & Yes & No & Arm & Superbubble & A \\
23 & 14:02:51.56 & +54:29:08.17 & 1.9 & 70 & 2.2E$-$15 & 0 & 0 & No & No & Interarm & Type Ia & B \\
25 & 14:02:53.56 & +54:14:24.33 & 1.6 & 60 & 3.7E$-$15 & ... & ... & Yes\tablenotemark{e} & No & Arm & Core-Collapse & A \\
26 & 14:02:54.43 & +54:23:24.75 & 0.7 & 30 & 5.9E$-$16 & 0 & 2 & No & No & Arm & Type Ia & A \\
27 & 14:02:55.25 & +54:24:17.25 & 1.0 & 40 & 3.4E$-$16 & 0 & 4 & No & PL & Interarm & Core-Collapse & A \\
28 & 14:02:56.06 & +54:14:57.30 & 2.5 & 90 & 3.1E$-$15 & ... & ... & Yes\tablenotemark{e} & No & Arm & Core-Collapse & C \\
30 & 14:02:59.12 & +54:19:50.10 & 1.2 & 40 & 1.5E$-$15 & 0 & 0 & No & Th & Interarm & Type Ia & A \\
32 & 14:02:59.49 & +54:22:45.20 & 1.0 & 40 & 2.2E$-$15 & 0 & 1 & No & Am & Arm & Core-Collapse & C \\
33 & 14:03:00.49 & +54:20:02.36 & 0.8 & 30 & 2.7E$-$15 & 0 & 0 & No & Th & Arm & Uncertain & D \\
34 & 14:03:02.09 & +54:23:24.54 & 2.0$\times$1.3 & 70$\times$50 & 2.8E$-$15 & 0 & 4 & No & Th & Arm & Core-Collapse & A \\
40 & 14:03:09.36 & +54:18:31.84 & 1.0 & 40 & 2.5E$-$15 & 1+ & 1+ & Yes & Am & Interarm & Core-Collapse & A \\
41 & 14:03:10.56 & +54:23:27.11 & 0.8 & 30 & 1.3E$-$15 & 0+ & 11+ & Yes & No & Arm & Core-Collapse & A \\
42 & 14:03:11.05 & +54:22:03.49 & 1.2 & 40 & 4.7E$-$16 & 0 & 0 & No & PL & Interarm & Type Ia & A \\
43 & 14:03:12.22 & +54:18:52.23 & 1.7 & 60 & 2.4E$-$15 & 0 & 0 & No & Am & Interarm & Type Ia & A \\
44 & 14:03:12.45 & +54:23:27.28 & 1.1 & 40 & 5.8E$-$15 & 0 & 8 & No & Th & Arm & Core-Collapse & A \\
45 & 14:03:12.76 & +54:17:35.32 & 4.5$\times$2.7 & 160$\times$100 & 1.2E$-$14 & 4 & 10 & No & No & Arm & OB/H\,{\sc II} & A \\
46 & 14:03:12.72 & +54:19:00.78 & 1.5 & 50 & 1.4E$-$15 & 0 & 0 & No & Th & Interarm & Type Ia & A \\
47 & 14:03:13.02 & +54:24:39.13 & 2.7 & 100 & 2.0E$-$15 & 2 & 3 & No & No & Interarm & Superbubble & A \\
48 & 14:03:13.24 & +54:17:06.77 & 3.5$\times$2.9 & 130$\times$100 & 5.5E$-$15 & 5+ & 10+ & Yes & Am & Interarm & OB/H\,{\sc II} & A \\
49 & 14:03:13.29 & +54:21:56.74 & 0.5 & 20 & 1.4E$-$16 & 0 & 3 & No & Th & Inner galaxy & Core-Collapse & A \\
50 & 14:03:14.62 & +54:21:51.58 & 0.5 & 20 & 1.0E$-$15 & 0 & 2 & No & Th & Inner galaxy & Core-Collapse & A \\
51 & 14:03:17.31 & +54:17:10.58 & 3.1 & 110 & 8.5E$-$15 & 6+ & 15+ & Yes & No & Arm & Superbubble & A \\
52 & 14:03:18.02 & +54:17:54.08 & 0.7 & 30 & 3.8E$-$16 & 0 & 5 & No & No & Interarm & Core-Collapse & A \\
53 & 14:03:20.57 & +54:16:51.68 & 5.8$\times$8.6 & 210$\times$310 & 8.6E$-$15 & 22+ & 27+ & Yes & No & Arm & Superbubble & A \\
54 & 14:03:20.80 & +54:19:42.23 & 1.2 & 40 & 5.9E$-$15 & 2+ & 5+ & Yes & Th & Interarm & Core-Collapse & A \\
57 & 14:03:24.31 & +54:19:39.62 & 0.8 & 30 & 8.2E$-$16 & 0 & 0 & No & Am & Arm & Type Ia & A \\
58 & 14:03:24.56 & +54:25:00.83 & 3.0$\times$5.8 & 110$\times$210 & 6.7E$-$15 & 1 & 5 & No & No & Arm & Superbubble & A \\
59 & 14:03:24.84 & +54:17:21.37 & 1.7 & 60 & 1.1E$-$15 & 0 & 2 & No & No & Arm & Core-Collapse & A \\
60 & 14:03:25.42 & +54:18:22.64 & 1.0 & 40 & 2.1E$-$15 & 0 & 2 & No & No & Arm & Core-Collapse & B \\
62 & 14:03:26.38 & +54:24:33.22 & 2.2 & 80 & 7.3E$-$15 & 1+ & 6+ & Yes & No & Arm & Core-Collapse & B \\
65 & 14:03:27.32 & +54:18:31.36 & 0.8 & 30 & 4.2E$-$15 & 0+ & 1+ & Yes & Am & Arm & Core-Collapse & A \\
67 & 14:03:27.79 & +54:24:30.42 & 5.1$\times$2.5 & 180$\times$90 & 1.0E$-$14 & 0+ & 0+ & Yes & No & Arm & OB/H\,{\sc II} & C \\
69 & 14:03:28.54 & +54:24:11.21 & 3.4 & 120 & 1.1E$-$14 & 3+ & 3+ & Yes & No & Arm & OB/H\,{\sc II} & C \\
73 & 14:03:31.05 & +54:24:42.25 & 4.9$\times$1.8 & 180$\times$60 & 9.0E$-$15 & 5+ & 8+ & Yes & No & Arm & Superbubble & A \\
75 & 14:03:32.40 & +54:17:40.77 & 3.4 & 120 & 4.8E$-$15 & 8+ & 11+ & Yes & No & Arm & Superbubble & A \\
76 & 14:03:32.79 & +54:17:40.61 & 1.8 & 60 & 2.0E$-$15 & 1 & 7 & No & No & Arm & Core-Collapse & A \\
78 & 14:03:33.81 & +54:17:43.22 & 7.8 & 280 & 6.9E$-$15 & 20 & 26 & No & No & Arm & OB/H\,{\sc II} & C \\
81 & 14:03:35.63 & +54:18:12.51 & 0.9 & 30 & 8.2E$-$16 & 0 & 1 & No & Am & Interarm & Type Ia & A \\
82 & 14:03:35.98 & +54:27:16.62 & 5.6$\times$3.0 & 200$\times$110 & 4.3E$-$15 & ... & ... & ... & No & Interarm & Superbubble & C \\
83 & 14:03:36.05 & +54:19:23.26 & 9.3 & 330 & 2.3E$-$14 & 2+ & 2+ & Yes & Am & Interarm & Superbubble & A \\
85 & 14:03:40.21 & +54:18:20.52 & 1.7 & 60 & 1.2E$-$15 & 1 & 2 & No & No & Interarm & Diffuse H\,{\sc II} & C \\
88 & 14:03:51.78 & +54:21:03.51 & 3.5 & 130 & 3.3E$-$15 & 8+ & 13+ & Yes & No & Arm & Core-Collapse & A \\
89 & 14:03:52.46 & +54:21:30.51 & 2.3$\times$1.3 & 80$\times$50 & 4.4E$-$15 & 3+ & 33+ & Yes & No & Arm & OB/H\,{\sc II} & A \\
90\tablenotemark{f} & 14:03:52.88 & +54:21:17.31 & 5.0$\times$2.2 & 180$\times$80 & 6.9E$-$15 & 6 & 6 & No & No & Arm & OB/H\,{\sc II} & A \\
91 & 14:03:53.81 & +54:21:24.21 & 2.1 & 80 & 3.1E$-$15 & 1 & 5 & No & No & Arm & OB/H\,{\sc II} & C \\
\enddata
\tablenotetext{a}{Number of massive stars within the boundary of the SNR candidate.}
\tablenotetext{b}{Number of massive stars within 100 pc of the center of smaller SNR candidates or 170 pc for large SNR candidates with diameters greater than 200 pc.}
\tablenotetext{c}{X-ray sources are identified as either thermal (Th), power-law (PL), or ambiguous (Am).}
\tablenotetext{d}{Confidence of our classification of the object ranked using a letter grade system, with A being most confident and D being least confident.}
\tablenotetext{e}{Identified in F606W images only.}
\tablenotetext{f}{Unclear as to whether the SNR candidate is the entire nebula we identified or only the east edge.}
\end{deluxetable}

\clearpage

\begin{deluxetable}{cccc}
\tabletypesize{\scriptsize}
\tablewidth{0pc}
\tablecolumns{4}
\tablecaption{Expansion velocities of SNR candidates in M101}
\tablehead{
\colhead{MF Number} & \colhead{$V_{\rm sys}$ (\kms)\tablenotemark{a}} & \colhead{$\Delta$$V_{\rm blue}$ (\kms)} & \colhead{$\Delta$$V_{\rm red}$ (\kms)}}
\startdata
19 & 227 & $-$70 & $<$44  \\
24 & 181 & $-$161 & 189 \\
25 & 186 & $-$140 & 194 \\
27 & 263 & $-$59 & 88 \\
28 & 193 & $-$73 & 46 \\
32 & 263 & $-$119 & 101 \\
45 & 206 & $-$90 & 99 \\
48 & 197 & $-$119 & 80 \\
51 & 201 & $-$79 & $<$64 \\
52 & 204 & $-$62 & 46 \\
53 & 200 & $-$45 & 46 \\
59 & 209 & $-$71 & 43 \\
75 & 229 & $-$48 & 62 \\
78 & 229 & $-$71 & $<$48 \\
85 & 253 & $-$48 & $<$34 \\
89 & 262 & $-$46 & 63 \\
90 & 263 & $-$39 & 69 \\
91 & 264 & $-$50 & 80 \\
\enddata
\tablenotetext{a}{All values are in the heliocentric frame.}
\end{deluxetable}

\clearpage

%\begin{deluxetable}{cc}
%\tabletypesize{\scriptsize}
%\tablewidth{0pc}
%\tablecolumns{2}
%\tablecaption{M101 Regions}
%\tablehead{
%\colhead{Region} & \colhead{OB Stars 100 pc$^{-2}$}}
%\startdata
%IARM-1 & 0 - .012 \\
%IARM-2 & 0 - .009 \\
%IARM-3 & 0 - .021 \\
%IARM-4 & 0 - .025 \\
%IARM-5 & 0 - .006 \\
%IARM-6 & 0 - .015 \\
%IARM-7 & 0 - .020 \\
%\enddata
%\end{deluxetable}

%\clearpage

%\begin{deluxetable}{cc}
%\tabletypesize{\scriptsize}
%\tablewidth{0pc}
%\tablecolumns{2}
%\tablecaption{SNRs with only X-ray data}
%\tablehead{
%\colhead{MF Number} & \colhead{X-ray}}
%\startdata
%02 & No \\
%03 & No \\
%04 & No \\
%05 & No \\
%06 & No \\
%10 & No \\
%14 & No \\
%18 & No \\
%20 & Yes \\
%21 & Yes \\
%22 & Yes \\
%24 & Yes \\
%29 & Yes \\
%31 & No \\
%35 & No \\
%36 & No \\
%37 & Yes \\
%38 & No \\
%39 & No \\
%55 & No \\
%56 & No \\
%61 & Yes \\
%63 & No \\
%64 & Yes \\
%66 & No \\
%68 & No \\
%70 & No \\
%71 & Yes \\
%72 & No \\
%74 & No \\
%77 & Yes \\
%79 & No \\
%80 & No \\
%84 & No \\
%86 & No \\
%87 & No \\
%92 & No \\
%93 & Yes \\
%\enddata
%\end{deluxetable}

%\clearpage

%\end{document}

\begin{figure}
\epsscale{1.0}
\plotone{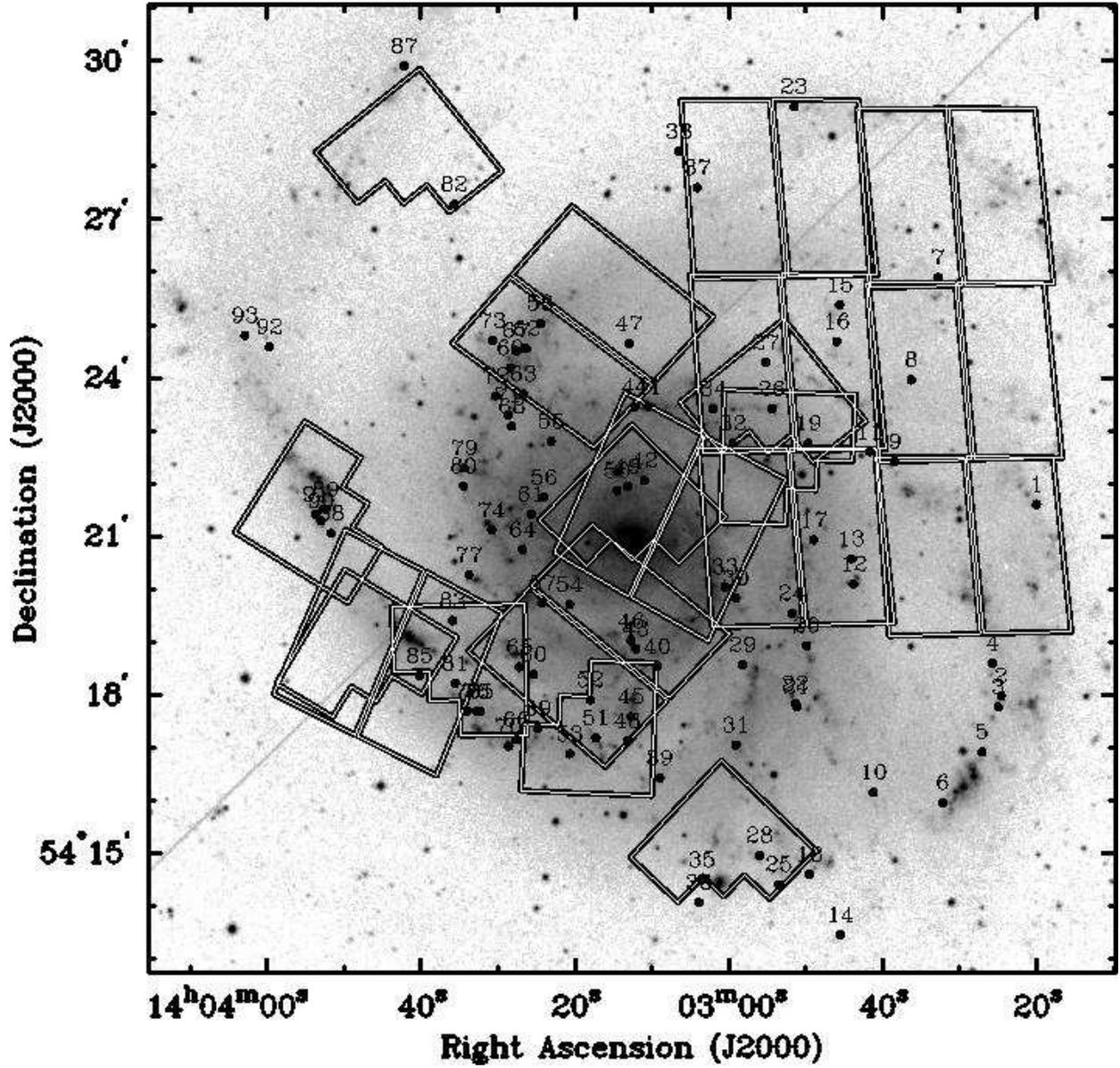}
\caption{Digitized Sky Survey red image of M101 overlaid with positions of archival \emph{HST} \ha\ observations and SNR candidates from \citet{MF97}.}
\end{figure}

\clearpage

\begin{figure}
\epsscale{1.0}
\plotone{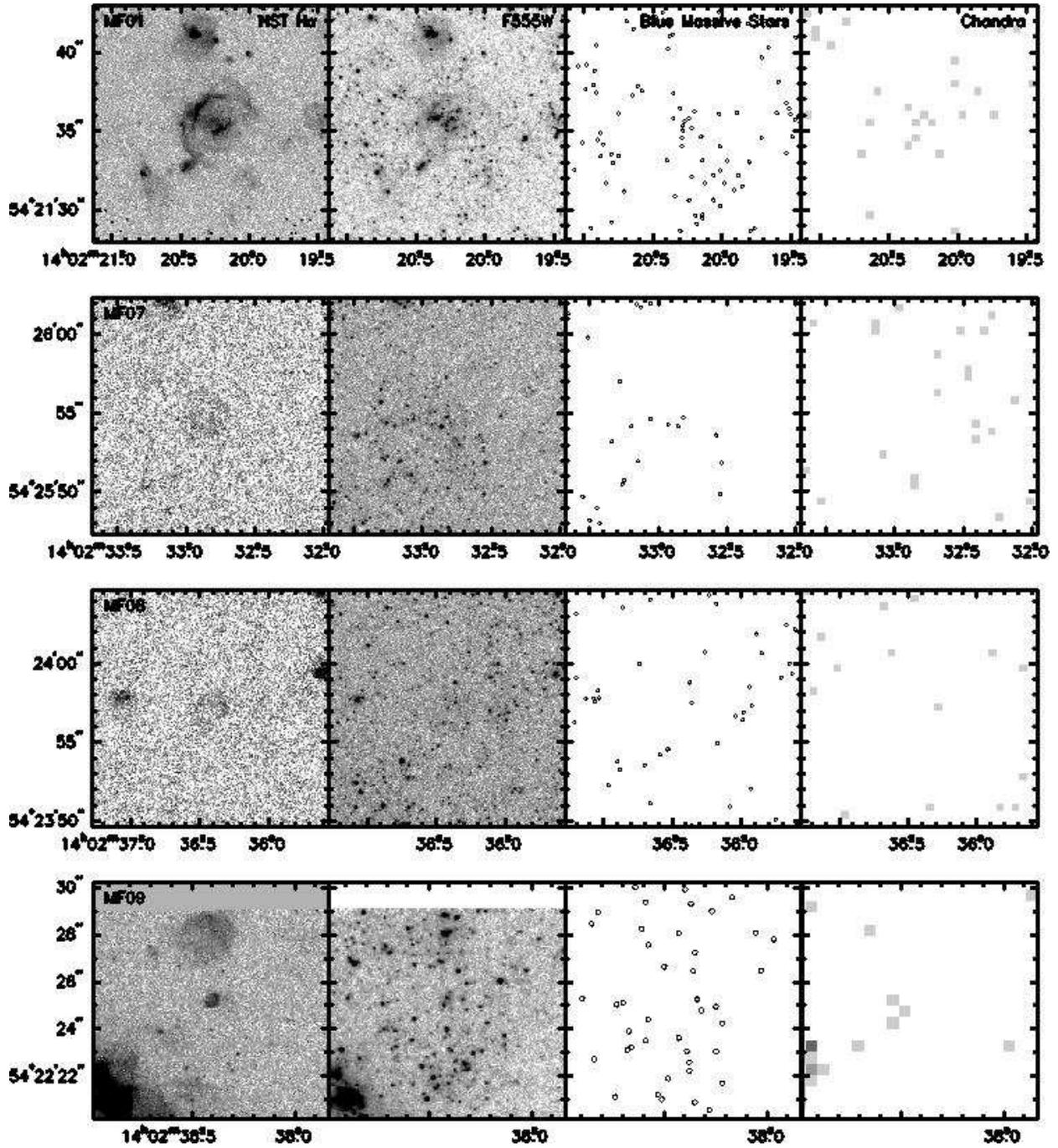}
\figurenum{2a}
%\end{figure}
%
%\clearpage
%
%\begin{figure}
%\epsscale{1.0}
%\plotone{figure2b.ps}
%\end{figure}
%
%\clearpage
%
%\begin{figure}
%\epsscale{1.0}
%\plotone{figure2c.ps}
%\end{figure}
%
%\clearpage
%
%\begin{figure}
%\epsscale{1.0}
%\plotone{figure2d.ps}
%
%\end{figure}
%
%\clearpage
%
%\begin{figure}
%\epsscale{1.0}
%\plotone{figure2e.ps}
%
%\end{figure}
%
%\clearpage
%
%\begin{figure}
%\epsscale{1.0}
%\plotone{figure2f.ps}
%\end{figure}
%
%\clearpage
%
%\begin{figure}
%\epsscale{1.0}
%\plotone{figure2g.ps}
%\end{figure}
%
%\clearpage
%
%\begin{figure}
%\epsscale{1.0}
%\plotone{figure2h.ps}
%\end{figure}
%
%\clearpage
%
%\begin{figure}
%\epsscale{1.0}
%\plotone{figure2i.ps}
%\end{figure}
%
%\clearpage
%
%\begin{figure}
%\epsscale{1.0}
%\plotone{figure2j.ps}
%\end{figure}
%
%\clearpage
%
%\begin{figure}
%\epsscale{1.0}
%\plotone{figure2k.ps}
%\end{figure}
%
%\clearpage
%
%\begin{figure}
%\epsscale{1.0}
%\plotone{figure2l.ps}
%\end{figure}
%
%\clearpage
%
%\begin{figure}
%\epsscale{1.0}
%\plotone{figure2m.ps}
%\end{figure}
%
%\clearpage

%\begin{figure}
%\epsscale{1.0}
%\plotone{figure2n.ps}
\caption{Images showing (from left to right) \ha\ emission, F555W emission, massive stellar population, and X-ray emission in the 0.5--2.0 keV band for the SNR candidates in this study.  Figures 2b-n are available in the electronic version of the journal.}
\end{figure}

\clearpage

\begin{figure}
\epsscale{1.0}
\figurenum{3}
\plotone{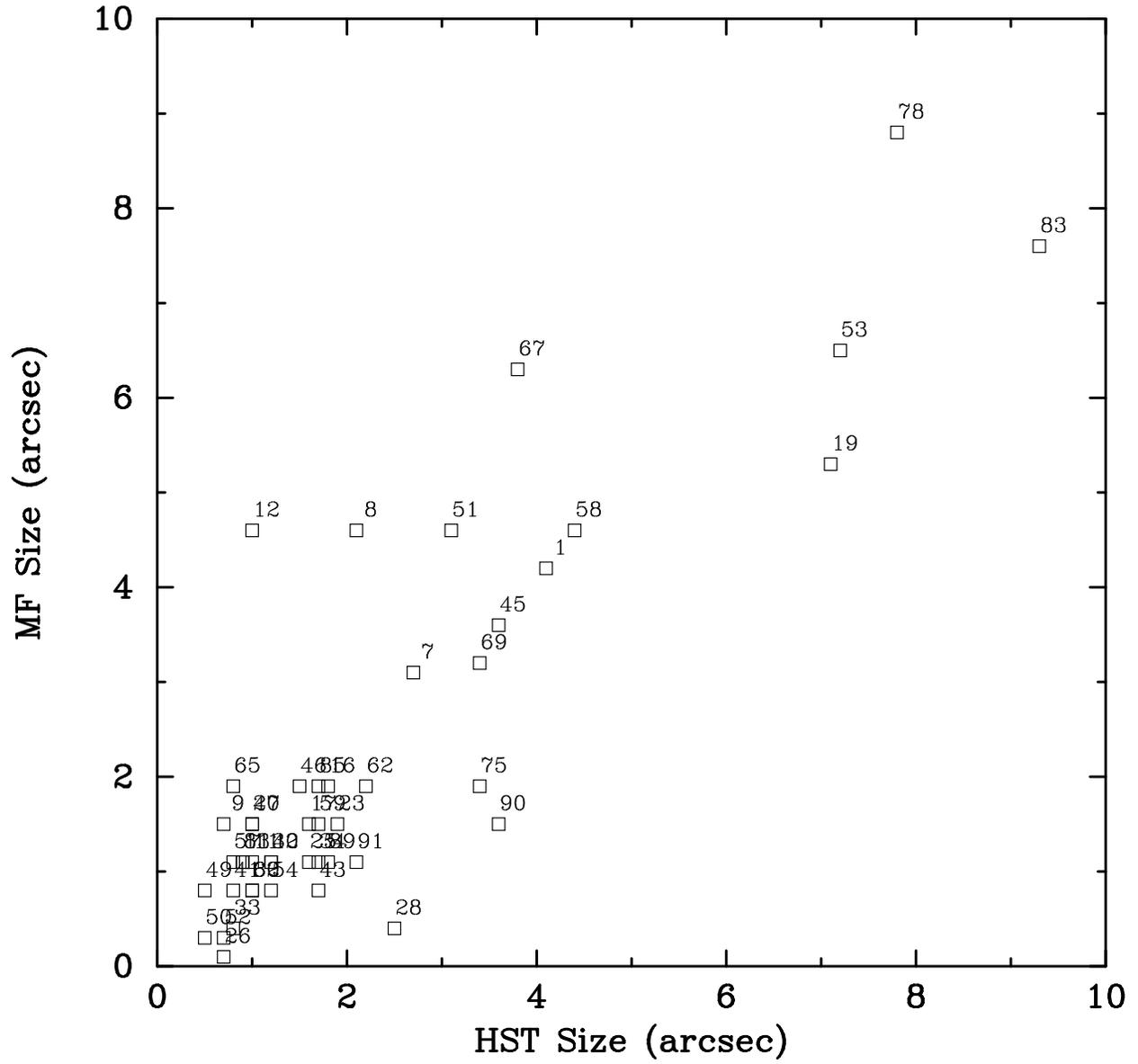}
\caption{Comparison of the linear sizes we measured, \emph{HST} size, to those measured by \citet{MF97}, MF size.}
\end{figure}

\clearpage

\begin{figure}
\epsscale{1.0}
\figurenum{4}
\plotone{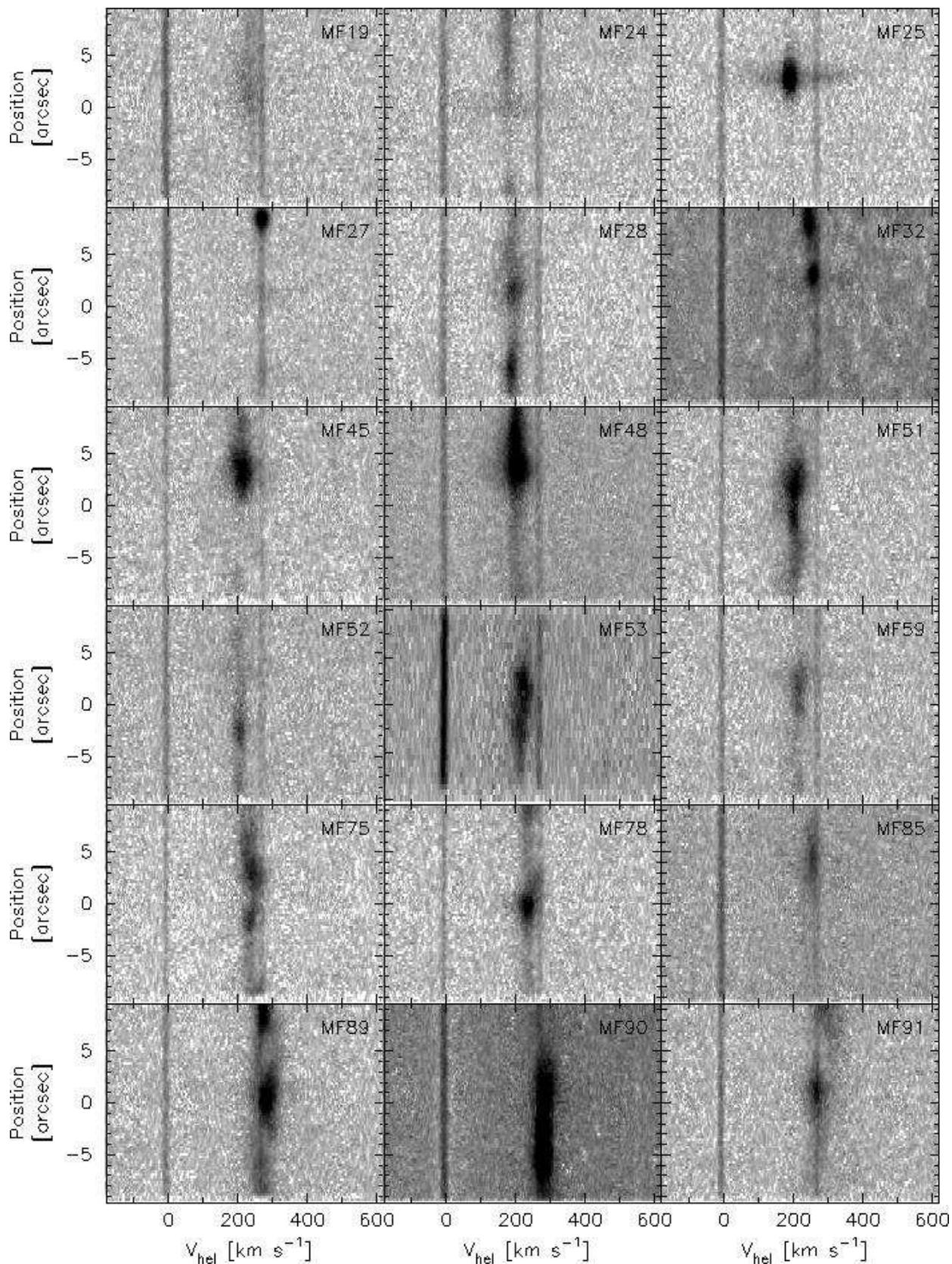}
\caption{\ha\ echellograms of 18 SNR candidates in M101. The horizontal axis is the heliocentric velocity, and the vertical axis is the position along the slit. The narrow lines at 0 and $\sim$270 
\kms\ correspond to the telluric \ha\ line and OH 6-1 P2 (3.5) lines, respectively. The name of the objects are marked in the upper right corner of each panel.}
\end{figure}

\clearpage

\begin{figure}
\epsscale{1.0}
\figurenum{5}
\plotone{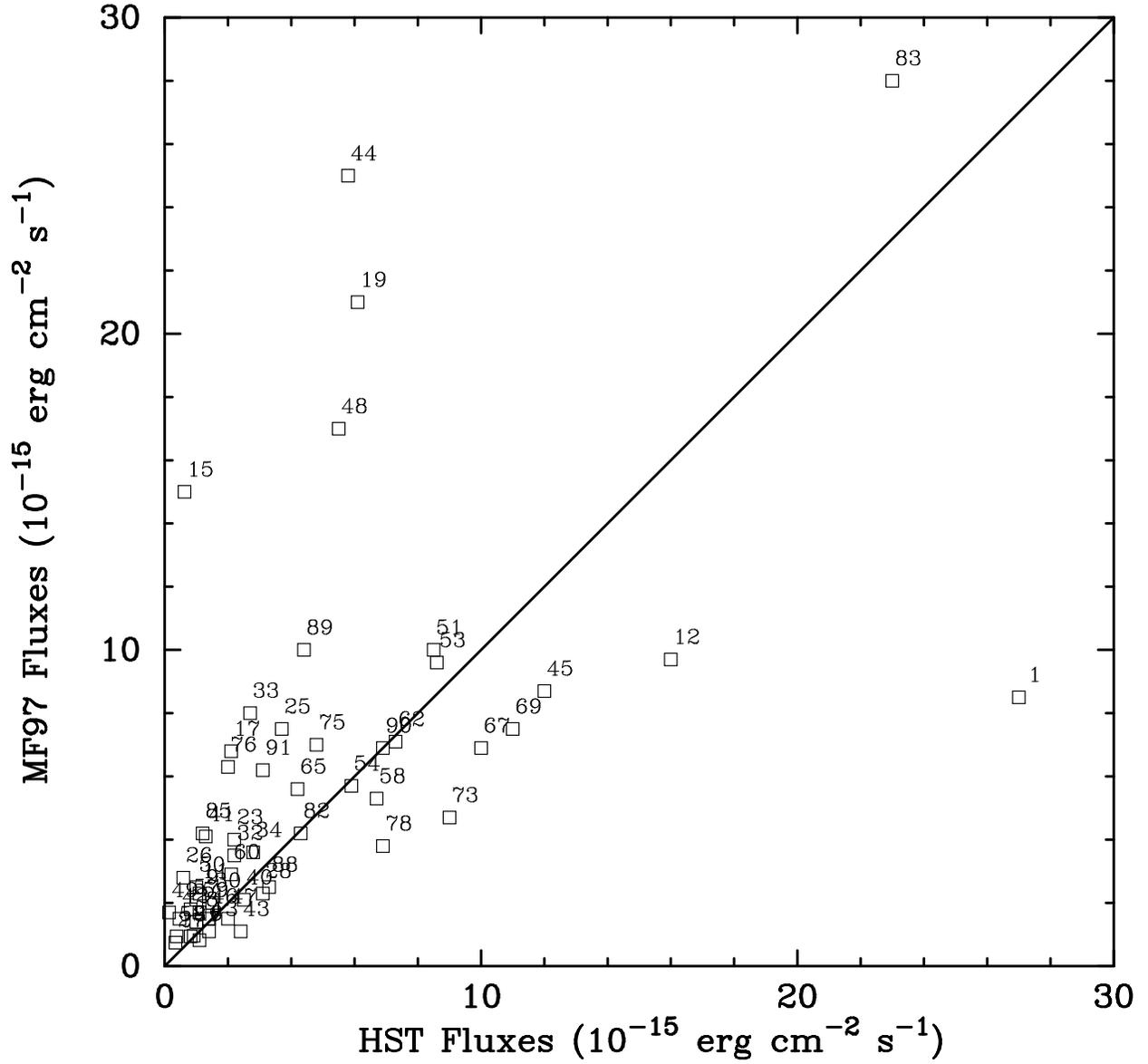}
\caption{Comparison of the \ha\ fluxes we measured to those measured by Matonick and Fesen.}
\end{figure}

\clearpage

\begin{figure}
\epsscale{1.0}
\figurenum{6}
\plotone{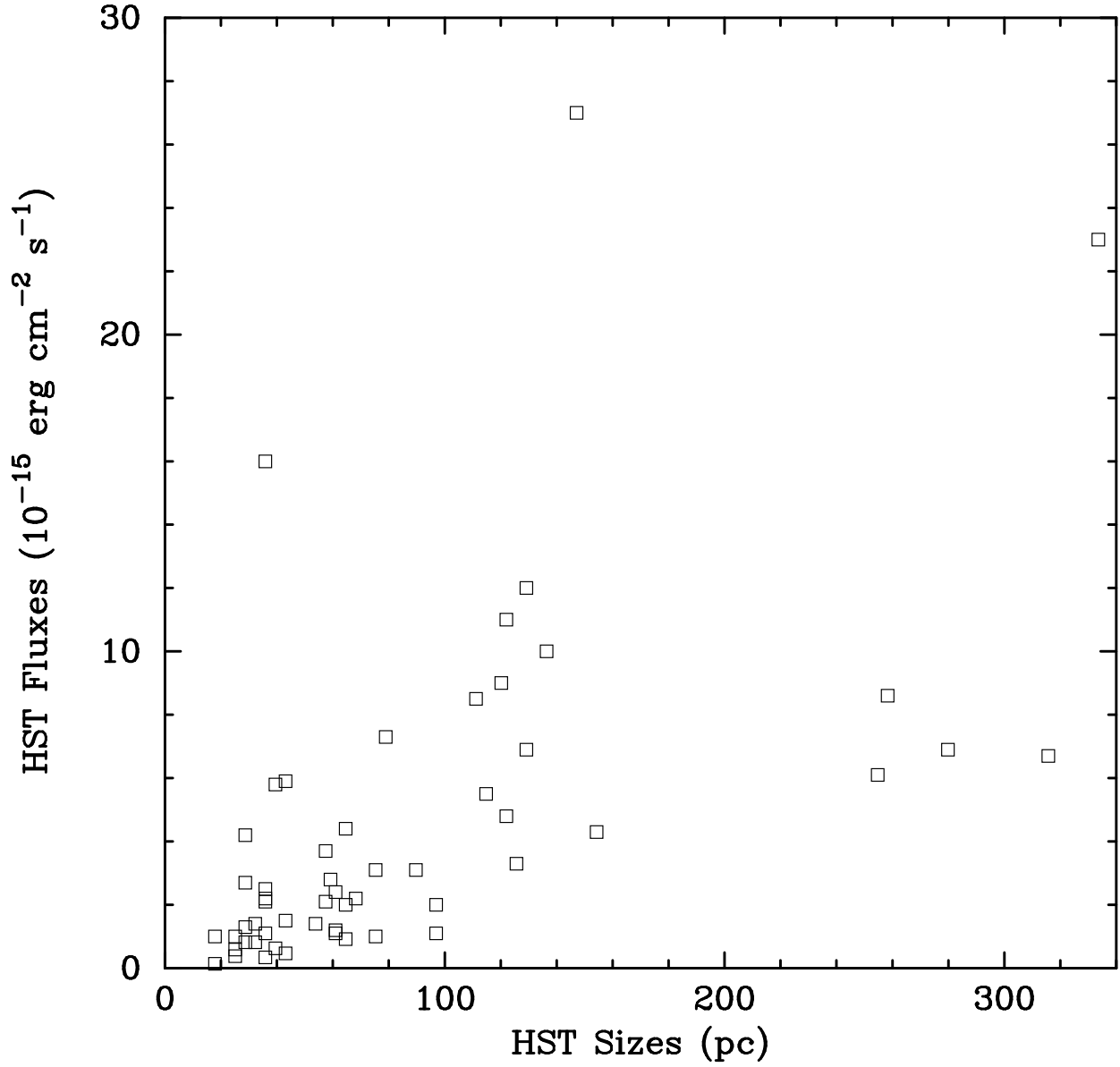}
\caption{Plot of size (diameter) of the SNR candidates versus \ha\ fluxes.}
\end{figure}

\clearpage

\begin{figure}
\epsscale{1.0}
\figurenum{7}
\plotone{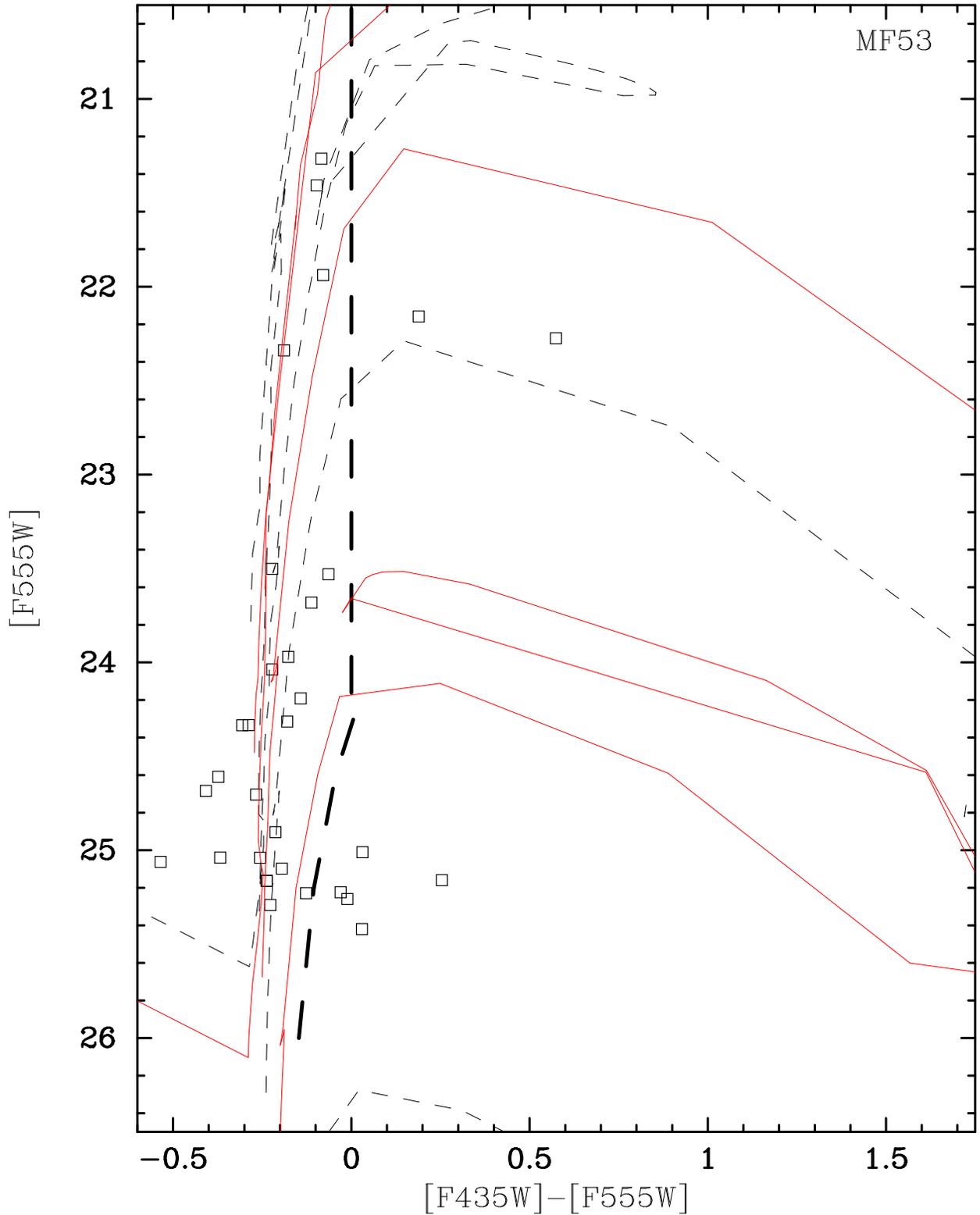}
\caption{CMD of the stars in the area around MF 53, shown as an example. Tracks from \citet{LS01} are shown with dashed black lines (5, 15, 25, and 60 $M_\odot$) and solid red lines (9, 20, and 40 $M_\odot$). The selection criteria for massive stars is shown as a heavy dashed line.}
\end{figure}

\clearpage

\end{document}